\documentclass[12pt]{article}
\pdfoutput=1
\usepackage{subfigure}
\usepackage{amssymb,amsmath}
\usepackage{graphicx}
\usepackage{color}
\usepackage{cancel}
\usepackage[colorlinks=true
,urlcolor=blue
,citecolor=blue
,linkcolor=blue
,pagecolor=blue
,linktocpage=true
,pdfproducer=medialab
]{hyperref}
\usepackage[a4paper,width=15.2cm]{geometry}

\usepackage{ulem}
\usepackage[section]{placeins}

\makeatletter \renewcommand{\@dotsep}{10000} \makeatother
\def\be{\begin{equation}}
\def\ee{\end{equation}}
\def\bea{\begin{eqnarray}}
\def\eea{\end{eqnarray}}
\def\bi{\begin{itemize}}
\def\ei{\end{itemize}}

\def\singone{\mathcal{X}_{1}}
\def\singtwo{\mathcal{X}_{2}}

\newcommand{\beq}{\begin{equation}}
\newcommand{\eeq}{\end{equation}}

\begin{document}
%Remove date before submitting to arXi

\date{\today}

\begin{center}
{\Large\bf 
Mass Spectrum and Higgs Profile in BLSSM } 
\end{center}

\begin{center}

{\Large
Cem Salih \"{U}n$^{a,b,}$\footnote{
Email: cemsalihun@uludag.edu.tr}
and
\"{O}zer \"{O}zdal$^{c,}$\footnote{
Email: ozerozdal@iyte.edu.tr}
}

\vspace{0.75cm}

{\it $^a$
Center of Fundamental Physics, Zewail City of Science and Technology, 6 October City, Cairo, Egypt

\hspace{-2.4cm}\it $^b$
Department of Physics, Uluda{\~g} University, TR16059, Bursa, Turkey
}\\
\vspace{0.2cm}
{\it  $^c$
Department of Physics, {\.I}zmir Institute of Technology, IZTECH, TR35430, {\. I}zmir, Turkey}

\vspace{1.5cm}
\section*{Abstract}
\end{center}
\noindent
We investigate the predictions on the mass spectrum and Higgs boson decays in the supersymmetric standard model extended by $U(1)_{B-L}$ symmetry (BLSSM). The model requires two singlet Higgs fields, which are responsible for the radiative breaking of $U(1)_{B-L}$ symmetry. It predicts degenerate right-handed neutrino masses ($1.7-2.2$ TeV) as well as the right-handed sneutrinos of mass $\lesssim 4$ TeV. The presence of right-handed neutrinos and sneutrinos trigger the baryon and lepton number violation processes, until they decouple from the Standard model particles. Besides, the model predicts rather heavy colored particles; $m_{\tilde{t}}, m_{\tilde{b}} \gtrsim 1.5$ TeV, while $m_{\tilde{\tau}} \gtrsim 100$ GeV and $m_{\tilde{\chi}_{1}^{\pm}} \gtrsim 600$ GeV. Even though, the implications are similar to minimal supersymmetric standard model, BLSSM can predict another Higgs boson lighter than 150 GeV. We find that the second Higgs boson can be degenerate with the lightest CP-even Higgs boson of mass about 125 GeV and contribute to the Higgs decay into two photons. In addition, it can provide an explanation for the excess in $h\rightarrow 4l$ at the mass scale $\sim 145$ GeV.

\newpage

%%%%%%%%%%%%%%%%%%%%%%%%%%%%%%%%%%%%%%%%%%%%%%%%%%%%%%%%%%%%
\renewcommand{\thefootnote}{\arabic{footnote}}
\setcounter{footnote}{0}

%%%%%%%%%%%%%%%%%%%%%%%%%%%%%%%%%%%%%%%%%%%%%%%%%%%%%%%%%%%%%

%\baselineskip 36pt
% Main body
%%%%%%%%%%%%%%%%%%%%%%%%%%
%\baselineskip 18pt
%%%%%%%%%%%%%%%%%%%%%%%%%%

%%%%%%%%%%%%%%%%%%%%%%%%%%%%
\section{Introduction}
\label{sec:intro}
%%%%%%%%%%%%%%%%%%%%%%%%%%%%

After the discovery of the Higgs boson of mass about $125-126$ GeV by the ATLAS \cite{:2012gk} and the CMS \cite{:2012gu} experiments, analyses have confirmed that the Standard Model (SM) predictions are in a very good agreement with the observations. Despite the fact that the SM has been completed with the Higgs boson discovery, there is no doubt that the SM is not a fundamental theory, since it is problematic in the Higgs boson due to the gauge hierarchy problem \cite{Gildener:1976ai} and the absolute stability of the Higgs potential \cite{hinstability}. However, experiments conducted at the Large Hadron Collider (LHC) have returned with no direct signal for new physics beyond the SM (BSM). In contrast, the experimental results almost overlap with the SM predictions. On the other hand, the Higgs boson may play a leading role in further analyses, since it provides a strong hints for BSM. In addition to the observed mass of the Higgs boson, detailed analyses have revealed some anomalies in decay channels of the Higgs boson. While combination of all decay channels excludes the range $\sim 150 < m_{h} < 1000$ GeV \cite{Khachatryan:2015cwa}, there is an excess in $h \rightarrow \gamma \gamma $ at $m_{\gamma\gamma}\approx 137$ GeV, in addition to that observed at $m_{\gamma\gamma} \approx 125$ GeV \cite{CMS:ril}. Similarly, $h\rightarrow 4l$ exhibits an excess at around $m_{4l}\approx 146$ GeV \cite{Chatrchyan:2013mxa}.

While one can count the SM Higgs boson for the observations at about 125 GeV, the anomalies at the higher scales can be considered as hints for the heavier SM-like Higgs bosons, which are not included in the SM. In this context, models with two or more Higgs bosons are worth studying in light of the anomalies mentioned above. Minimal supersymmetric extension of the SM (MSSM) is classified as a theory with two Higgs doublets and it is arguably one of the prime candidate for BSM, since it provides a resolution to the gauge hierarchy problem. The two Higgs doublets yield five physical Higgs boson states after electroweak symmetry breaking (EWSB), and they may offer a number of different scenarios to explain the anomalies in decay modes of the Higgs bosons \cite{Ferreira:2012nv}. 

Even though it is possible to fit the low scale implications with the observed data in the MSSM framework, one can also consider the high scale origin, since the three gauge couplings of the SM nicely unify at the grand unification scale ($M_{{\rm GUT}} \approx 2\times 10^{16}$ GeV). Stabilizing  the Higgs boson mass at all the energy scales, one can connect $M_{{\rm GUT}}$ to the electroweak scale ($M_{{\rm EW}}$) through the renormalization group equations (RGEs). In such models, a large number of low scale MSSM parameters can be calculated through RGEs with a few free parameters defined at $M_{{\rm GUT}}$. Although MSSM is compatible with the current experimental results, the Higgs boson results bring severe constraints on the sparticle spectrum. As is well-known, the tree-level Higgs boson mass prediction is inconsistently low in the MSSM, and hence, one needs to utilize the loop corrections in order to realize the observed Higgs boson mass. Since the first two families have negligible Yukawa couplings to the Higgs boson, the third family provides the main source for such contributions. The sbottom and stau contributions exhibit $\tan\beta$ enhancement, and they can easily destabilize the Higgs potential; hence, their contributions are strongly constrained by the vacuum stability which allows only minor contributions from the sbottom and stau sector \cite{Carena:2012mw}. On the other hand, the contribution from stop is proportional to $\cot\beta$ and it has more freedom to satisfy the vacuum stability. In this context, the Higgs boson mass can be fed with the loop contributions from the stop sector, and it constrains the stop mass to the multi-TeV range, or it requires rather large mixing between stops. Even though it is possible to realize the stop of mass about top quark mass in the presence of the large mixing, the parameter space needs to be highly fine-tuned in this case \cite{Demir:2014jqa}. The stop mass is bounded from below to a few hundred GeV if one imposes the fine-tuning condition \cite{Gogoladze:2013wva}. 

Besides the Higgs boson results, another severe constraint comes from the observation of the rare decay $B_{s}\rightarrow \mu^{+}\mu^{-}$ with the branching ratio ${\rm BR}(B_{s}\rightarrow \mu^{+}\mu^{-})=3.2^{+1.5}_{-1.2}\times 10^{-9}$ \cite{Aaij:2012nna}. This discovery is only another success of the SM, since its prediction for this rare decay more or less overlaps with the observation \cite{Bobeth:2013uxa}. The small window in the prediction for this rare decay severely constrains the models for BSM. In MSSM, the supersymmetric contributions to $B_{s}\rightarrow \mu^{+}\mu^{-}$ comes from the CP-odd Higgs boson exchange, which is proportional to $(\tan\beta)^{6}/m_{A}$, where $m_{A}$ is CP-odd Higgs boson mass. Accordingly, $m_{A}$ needs to be heavy enough to suppress the $\tan\beta$ enhancement which requires $m_{A} \gtrsim 500$ GeV \cite{Gogoladze:2014vea}. This constraint bounds the heavier CP-even Higgs boson mass ($m_{H}$) since $m_{H}\approx m_{A}$. After all, despite the abundance of the physical Higgs boson states, MSSM cannot fit them in the mass range $m \lesssim 150$ GeV consistently with the experimental results, especially when it is constrained from $M_{{\rm GUT}}$.

Considering the minimality the discussion above can be concluded that MSSM may not cover the full story and one may consider some extension of the MSSM gauge group. One of the simplest extensions is imposing an extra $U(1)$ symmetry. Such an extension can be obtained from an underlying GUT theory involving a gauge group larger than $SU(5)$ \cite{Langacker:2008yv}. Among the many different realizations of $G_{{\rm MSSM}}\times U(1)_{X}$, $U(1)_{B-L}$ provides a favorable framework, since the anomaly cancellation can be achieved by adding three MSSM singlets, and the right-handed neutrino is the first choice for such singlet fields. In this context, an anomaly free $U(1)_{B-L}$ extension of MSSM provides a natural framework for the established non-zero neutrino masses \cite{Wendell:2010md} through the seesaw mechanisms. Besides, the invariance under $U(1)_{B-L}$ gauge group also imposes the $R-$parity conservation which is assumed in the MSSM to avoid fast proton decay. Hence, $R-$parity violation can be constrained by the smallness of the neutrino masses. Moreover, $R-$parity conservation can be maintained if $U(1)_{B-L}$ symmetry is broken spontaneously \cite{Aulakh:1999cd}. Indeed, it was shown that $U(1)_{B-L}$ symmetry can be broken radiatively through a similar mechanism to the radiative electroweak symmetry breaking (REWSB) in the MSSM \cite{Khalil:2007dr}. One can introduce a field whose non-zero vacuum expectation value (VEV) breaks the $U(1)_{B-L}$ symmetry. Hence, this field should carry $B-L$ charge and it is preferably singlet under the MSSM gauge group. If its $B-L$ charge is 2, then the $R-$parity conservation can be maintained. The holomorphy condition of the superpotential requires another MSSM singlet field whose $B-L$ charge is $-2$ in order to write the invariant Lagrangian under $U(1)_{B-L}$. Hence the MSSM extended by $U(1)_{B-L}$ (BLSSM) proposes two singlet Higgs fields ($\singone$ and $\singtwo$ with $-2$ and $+2$ B-L charges respectively) which can be counted for the observed anomalies in the Higgs decays at the mass scales other than $\sim 125$ GeV.

The rest of the paper is organized as follows: In Section \ref{sec:model} we briefly describe the model with an emphasize on the Higgs sector. After we summarize the scanning procedure and the experimental constraints employed in our analysis in Section \ref{sec:scan}, we present our results for the mass spectrum in Section \ref{sec:results}. We also briefly mention about Leptogenesis in this section. In Section \ref{subsec41} we consider the Higgs boson decays into two photons and four leptons. Finally we summarize and conclude in Section \ref{sec:conc}.

\section{Model Description}
\label{sec:model}

In this section, we review the BLSSM model with an emphasize to its Higgs sector. The superpotential in this model is given by

\begin{equation*}
W=\mu H_{u}H_{d}+Y_{u}^{ij}Q_{i}H_{u}u^{c}_{j}+Y_{d}^{ij}Q_{i}H_{d}d^{c}_{j}+Y_{e}^{ij}L_{i}H_{d}e^{c}_{j}
\end{equation*}
\begin{equation}\hspace{-1.0cm}
+Y_{\nu}^{ij}L_{i}H_{u}N^{c}_{i}+ Y^{ij}_{N}N^{c}_{i}N^{c}_{j}\singone+\mu^{\prime}\singone\singtwo
\label{superpotential}
\end{equation}
where the first line of Eq.(\ref{superpotential}) is the usual terms of the MSSM, while the second line includes the additional interactions from the right-handed neutrino $N^{c}_{i}$, and the singlet Higgs fields $\singone$, $\singtwo$ with $-2$ and $+2$ $B-L$ charges respectively. Once the model includes the right-handed neutrino, then one can add a Yukawa interaction term for the neutrinos, and $Y_{\nu}$ stands for the Yukawa coupling for the neutrinos. Similarly, $Y_{N}$ is the Yukawa coupling between $N^{c}_{i}$ and $\singone$. Finally $\mu^{\prime}-$term is the bilinear mixing between $\singone$ and $\singtwo$. The relevant soft supersymmetry breaking (SSB) Lagrangian is

\begin{equation*}
-\mathcal{L}_{\cancel{{\rm SUSY}}}=-\mathcal{L}^{{\rm MSSM}}_{\cancel{{\rm SUSY}}} +m^{2}_{\tilde{N}^{c}}|\tilde{N}^{c}|^{2} + m_{\singone}^{2}|\singone|^{2}+ m_{\singtwo}^{2}|\singtwo|^{2}
\end{equation*}
\begin{equation}\hspace{-1.0cm}
+A_{\nu}\tilde{L}H_{u}\tilde{N}^{c} +A_{N}\tilde{N}^{c}\tilde{N}^{c}\singone 
\label{SSBLag}
\end{equation}
\begin{equation*}
+\frac{1}{2}M_{\tilde{B}^{\prime}}\tilde{B}^{\prime}\tilde{B}^{\prime}+B(\mu^{\prime}\singone\singtwo+ {\rm h.c.})
\end{equation*}
where $\mathcal{L}^{{\rm MSSM}}_{\cancel{{\rm SUSY}}}$ includes the SSB terms of MSSM, while the rest associated with the $B-L$ symmetry. The meaning of the terms are similar to those in the MSSM. $m_{\tilde{N}^{c}}$, $m_{\singone}$ and $m_{\singtwo}$ are the SSB mass terms for the right-handed sneutrino, $\singone$ and $\singtwo$, while $A_{\nu}$ and $A_{N}$ are the trilinear scalar interaction terms between the neutrinos and MSSM Higgs doublets and BLSSM Higgs singlets respectively. $M_{\tilde{B}^{\prime}}$ is the SSM mass term for the gaugino $\tilde{B}^{\prime}$ associated with the $B-L$ gauge group. Note that there exists a vector-boson partner $Z^{\prime}$ whose mass is severely constrained by the current experimental results ($m_{Z^{\prime}} \gtrsim 2.5$ TeV). 

Note that, in contrast to its non-SUSY version, BLSSM does not allow mixing between the doublet and singlet Higgs fields through the superpotential and SSB Lagrangian. Therefore, the Higgs potential for these fields do not couple each other. Then, the singlet Higgs potential can be written as

\begin{equation*}
V(\singone , \singtwo)=\mu^{\prime 2}_{1}|\singone|^{2}+\mu^{\prime 2}_{2}|\singtwo|^{2}-\mu^{\prime}_{3}(\singone \singtwo + {\rm h.c.})
\end{equation*}
\begin{equation}
+\frac{1}{2}g_{BL}^{2}(|\singone|^{2}-|\singtwo|^{2})^{2}
\label{singpotential}
\end{equation}  
where $\mu^{\prime 2}_{1}=m_{\singone}^{2}+\mu^{\prime 2}$, $\mu^{\prime 2}_{1}=m_{\singtwo}^{2}+\mu^{\prime 2}$, $\mu^{\prime}_{3}=-B\mu^{\prime}$ and $g_{BL}^{2}$ is the gauge coupling associated with the $B-L$ gauge group. Since the potential in Eq.(\ref{singpotential}) is in the same form as the MSSM Higgs potential, its minimization yields similar relations regarding to the spontaneous breaking of $U(1)_{B-L}$ symmetry and the stability of the vacuum. When $m_{\singone}$ or $m_{\singtwo}$ (or both) is negative, the vacuum corresponds to non-zero vacuum expectation values (VEVs) $v'_{\singone}=\langle \singone \rangle$ and $v'_{\singtwo}=\langle \singtwo \rangle$.

A similar analysis in the radiative electroweak symmetry breaking (REWSB) can hold also for the $B-L$ symmetry breaking.  The coupling $Y_{N}$ between the right-handed neutrinos and $\singone$ negatively contributes to $m_{\singone}^{2}$, and if it is large enough, $m_{\singone}^{2}$ can turn to be negative from some positive values and it triggers the spontaneous $B-L$ symmetry breaking. It should be noted here that the interaction term between the right-handed neutrinos and $\singone$ induces a Majorana mass term $-Y_{N}v'_{\singone}\tilde{N}^{c}\tilde{N}^{c}$, which can destabilize the vacuum. For large values of $Y_{N}$, the global minimum can corresponds to non-zero  VEV of the right-handed sneutrinos, and hence it breaks the $R-$parity \cite{MOLINA:2014uha}. Hence, $Y_{N}$ should be large enough to trigger the spontaneous $B-L$ symmetry breaking, and small enough to preserve the $R-$parity. 

\begin{figure}[ht!]
\centering
\includegraphics[scale=0.30]{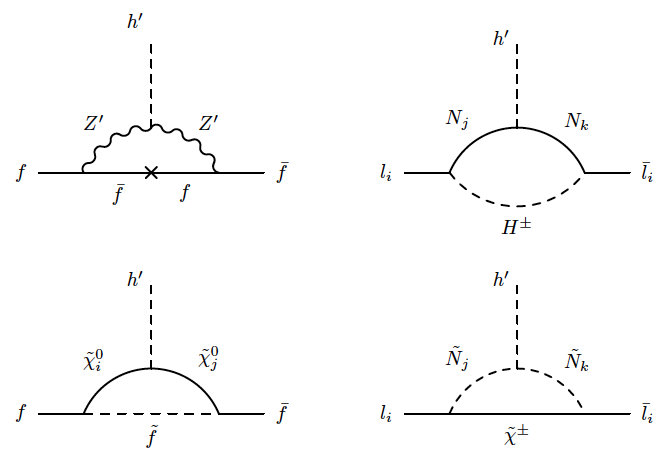}
\caption{The effective Yukawa interactions between the singlet Higgs and fermions. The top diagrams illustrate the non-SUSY loops, while the bottom diagrams displays the SUSY interference.}
\label{fig1}
\end{figure}

The spontaneous symmetry breaking mixes the fields and yields non-diagonal mass matrices. Since the two Higgs sector are not coupled to each other, their mass square matrices can be diagonalized independently, and hence the mass eigenstates related to $\singone$ and $\singtwo$ remain singlet under the SM gauge group, but they can still participate in the interactions with the MSSM fields through loops. Figure \ref{fig1} illustrates the effective Yukawa interactions between the singlet Higgs boson and matter particles. The top diagrams show the non-SUSY sector, while the bottom diagrams display the SUSY interference, since $\tilde{f}$, $\tilde{N}$, $\tilde{\chi}^{0}$ and $\tilde{\chi}^{\pm}$ stand for the sfermions, right-handed sneutrinos, neutralinos and charginos respectively. Since we assume that there is no mixing between the doublet and singlet Higgs fields, the singlet Higgs fields do not interact  with the left-handed neutrinos. The top left diagram includes a $Z'$ loop, and it is more likely suppressed due to the heavy mass bound on $Z'$. The contributions from the right top and bottom diagrams depend on the mixing in the neutrino sector and $Y_{N}$. The supersymetric contributions depend also on the sparticle masses running in the loops. The sneutrino loop is probably suppressed, since the sneutrino masses are at the order of TeV scale. The neutralino loop can lead to interesting results. Indeed, the contribution from the bottom left diagram depends on the $\tilde{B}'$ mass, which mixes with other neutralinos. Since there is no specific mass bound on $\tilde{B}'$, it can be as light as about 100 GeV, and it can even form the lightest neutralino \cite{Khalil:2015wua}.

Even though the discussion above shows that the singlet Higgs field can still alter the low scale phenomenology, it is rather a naive discussion, since the mixing between the Higgs fields is assumed not to be generated from another source or induced by the loop corrections. However, the invariance  principle allows the Lagrangian to include a cross term between the strength tensors of gauge fields associated with the $U(1)$ gauge groups, $-\kappa_{ab} B^{a}_{\mu\nu}B^{b,\mu\nu}$, where $B_{\mu\nu}$ is the field strength tensor of a $U(1)$ gauge field, $a,b = Y,~B-L$, the hypercharge and $B-L$ charge respectively, $\kappa_{ab}$ is an anti-symmetric tensor which includes the mixing of $U(1)_{a}$ and $U(1)_{b}$ gauge fields. This mixing couples the $B-L$ sector to the MSSM sector, and even if it is set to zero at $M_{{\rm GUT}}$, it can be induced through RGEs \cite{Holdom:1985ag}. In the case of non-zero gauge kinetic mixing, the gauge covariant derivative takes a non-canonical form as

\begin{equation}
\mathcal{D}_{\mu}=\partial_{\mu}-i(Y,B-L)\left(\begin{array}{cc} g_{Y} & \tilde{g} \\ \tilde{g}' & g_{B-L} \end{array}\right)\left(\begin{array}{c} B_{\mu} \\ B'_{\mu}\end{array}\right)
\label{noncanonD}
\end{equation}
where we have expressed the field in the flavor basis. Following the discussion in \cite{O'Leary:2011yq}, we will consider a basis by rotating the fields such that 

\begin{equation*}
\left(\begin{array}{cc} g_{Y} & \tilde{g} \\ \tilde{g}' & g_{B-L} \end{array}\right)\rightarrow \left(\begin{array}{cc} g_{1} & g_{YB} \\ 0 & g_{4} \end{array}\right)
\end{equation*}
where $g_{1}$ corresponds to the measured hypercharge coupling which is modified in BLSSM as given along with $g_{4}$ and $g_{YB}$ in \cite{O'Leary:2011yq,Chankowski:2006jk}. 

With non-zero mixing between the $U(1)$ gauge fields, a contribution from $Z-$boson loop similar to the top left diagram in Figure \ref{fig1} exists. On the other hand, the  gauge kinetic mixing affects the mixing in the other sectors. Especially it induces a tree-level mixing between the MSSB doublet Higgs and BLSSM Higgs fields proportional to $g_{YB}$. As a consequence of non-zero gauge kinetic mixing, the two Higgs sector becomes coupled and their mass square matrices should be diagonalized together. Then all the mass eigenstates couple to the MSSM particles at tree-level. In this case the contributions in Figure \ref{fig1} represent the corrections to the tree-level couplings. Note that a non-zero mixing in the Higgs sector brings also contributions from the chargino loop. Having these extra Higgs bosons coupled to the MSSM particles leads to contributions to the observed processes associated with the Higgs sector. Interestingly these Higgs bosons can be counted for the excesses observed in the higher mass scales mentioned in the previous section.

Before concluding this section, we should comment about the right-handed neutrino contributions to the Higgs bosons. As seen from Eq.(\ref{superpotential}), the presence of the right-handed neutrino allows to have a Yukawa interaction term involving with $H_{u}$. This term yields contributions to the SM-like Higgs boson in addition to the stop sector, and it may loose the mass bound on the stops and consequently improves the fine-tuning in the model. However, after the electroweak symmetry breaking, this term induces a Dirac mass for the neutrinos. Smallness of the established neutrino masses strictly bounds the associated Yukawa coupling to very small ranges ($Y_{\nu} \lesssim 10^{-7}$) \cite{Abbas:2007ag}, which strongly suppresses the contributions to the Higgs boson from the neutrino sector. Therefore BLSSM and MSSM yield similar low scale phenomenology for the Higgs boson. One can adopt the inverse seesaw mechanism into BLSSM \cite{Mohapatra:1986bd}, which allows $Y_{\nu}$ to be at the order of unity. Hence, the contribution from the right-handed neutrino sector to the Higgs boson cannot be neglected \cite{Elsayed:2011de}. Besides, the singlet Higgs fields interact with another singlet field with non-zero $B-L$ charge along with the right-handed neutrino, which yields a significant contribution to masses of the extra Higgs bosons. Hence, in the presence of the inverse seesaw mechanism, it is not easy to fit at least one more Higgs to the scale $m_{h} \lesssim 150$ GeV, when the model is constrained from $M_{{\rm GUT}}$. In other words, seeking for the second Higgs boson of mass less than 150 GeV leads also to very light SM-like Higgs boson ($\ll 125$ GeV). Note that even in BLSSM without inverse seesaw, the right-handed neutrino sector is still effective on the singlet Higgs boson masses, but since the SM-like Higgs boson does not acquire significant contributions from right-handed neutrinos, the singlet Higgs boson mass can be found light without affecting the SM-like Higgs boson mass. The RGEs for the singlet Higgs fields and the right-handed neutrino from $M_{{\rm GUT}}$ to the low scale are \cite{Khalil:2007dr}

\begin{equation*}
\frac{dm_{\singone}^{2}}{dt}=\frac{1}{16\pi^{2}}\left[6g_{BL}M_{BL}^{2}-2Y_{N}(m^{2}_{\singone}+2m^{2}_{N}+A^{2}_{N})\right] \tag{5-a}
\label{RGEsingone}
\end{equation*}

\begin{equation*}\hspace{-5.2cm}
\frac{dm_{\singtwo}^{2}}{dt}=\frac{1}{16\pi^{2}}6g_{BL}M_{BL}^{2} \tag{5-b}
\label{RGEsingtwo}
\end{equation*}

\setcounter{equation}{5}

\begin{equation}
\frac{dm_{N}^{2}}{dt}=\frac{1}{16\pi^{2}}\left[\frac{3}{2}g_{BL}M_{BL}^{2} -Y_{N}(m^{2}_{\singone}+2m^{2}_{N}+A^{2}_{N}) \right]
\end{equation}

\section{Scanning Procedure and Experimental Constraints}
\label{sec:scan}

We have employed SPheno 3.3.3 package \cite{Porod:2003um} obtained with SARAH 4.5.8 \cite{Staub:2008uz}. In this package, the weak scale values of the gauge and Yukawa couplings presence in MSSM are evolved to the unification scale $M_{{\rm GUT}}$ via the renormalization group equations (RGEs). $M_{{\rm GUT}}$ is determined by the requirement of the gauge coupling unification through their RGE evolutions. Note that we do not strictly enforce the unification condition $g_{1}=g_{2}=g_{3}$ at $M_{{\rm GUT}}$ since a few percent deviation from the unification can be assigned to unknown GUT-scale threshold corrections \cite{Hisano:1992jj}. With the boundary conditions given at $M_{{\rm GUT}}$, all the SSB parameters along with the gauge and Yukawa couplings are evolved back to the weak scale. Note that the gauge coupling associated with the $B-L$ symmetry is determined by the unification condition at the GUT scale by imposing $g_{1}=g_{2}=g_{4}\approx g_{3}$.

The requirement of radiative electroweak symmetry breaking (REWSB) \cite{Ibanez:1982fr} puts an important theoretical constraint on the parameter space. In our case, also the radiative $B-L$ symmetry breaking are required, but this requirement constrains rather right-handed neutrino sector and their coupling $Y_{N}$ to the $B-L$ singlet Higgs fields. 

We have performed random scans over the following parameter space

\begin{equation}
\begin{array}{ccc}
0 \leq & m_{0} & \leq 3 ~{\rm (TeV)} \\
0 \leq & M_{1/2} & \leq 5 ~{\rm (TeV)} \\
1.2 \leq & \tan\beta & 60 \\
-3 \leq & A_{0}/m_{0} & \leq 3 \\
\mu > 0, & \mu^{\prime}>0, & m_{t}=173.3~{\rm GeV}
\end{array}
\label{paramSP}
\end{equation}  
where we restrict ourselves only the universal boundary conditions in which $m_{0}$ denotes the SSB mass term for all the scalar including the MSSM doublet and BLSSM singlet Higgs fields, while $M_{1/2}$ stands for the SSB mass terms for the gauginos including one associated with the $U(1)_{B-L}$ gauge group. $A_{0}$ is the SSB trilinear scalar interacting term, $\tan\beta$ is the ratio of VEVs of the MSSM Higgs doublets.Note that the ratio of VEV of the BLSSM singlet Higgs fields is, in principle, a free parameter. In our scan it is restricted to be approximately unity ($\tan\beta^{\prime} \equiv v_{\singone}/v_{\singtwo} \approx 1-1.2$). Besides, $\mu$ is the bilinear mixing of the MSSM doublet Higgs fields, while $\mu^{\prime}$ is of the BLSSM singlet Higgs fields. In addition, $m_{t}$ is the top quark mass and we set it to its central value \cite{Group:2009ad}. Note that the sparticle spectrum is not too sensitive to one or two sigma variation in the top quark mass \cite{Gogoladze:2011db}, but it can shift the Higgs boson mass by $1-2$ GeV \cite{Gogoladze:2011aa}. Finally, we also vary the couplings $Y_{\nu}$, $g_{YB}$ in the perturbative level and fix $Y_{N}\approx 0.4$. 

In scanning the parameter space, we use our interface which employs Metropolis-Hasting algorithm as described in \cite{Belanger:2009ti}. All the data points satisfy the requirement of REWSB. After collecting the data, we impose the mass bounds on all the particles \cite{Agashe:2014kda}, and the following phenomenological constraints:  

\begin{eqnarray} 
m_h  = 123-127~{\rm GeV}~~&\cite{:2012gk, :2012gu}& 
\\
m_{\tilde{g}} \geq 1.8~{\rm TeV} 
\\
m_{\tilde{\tau}} \geq 105~{\rm GeV}
\\
0.8\times 10^{-9} \leq{\rm BR}(B_s \rightarrow \mu^+ \mu^-) 
  \leq 6.2 \times10^{-9} \;(2\sigma)~~&\cite{Aaij:2012nna}& 
\\ 
2.99 \times 10^{-4} \leq 
  {\rm BR}(b \rightarrow s \gamma) 
  \leq 3.87 \times 10^{-4} \; (2\sigma)~~&\cite{Amhis:2012bh}&  
\\
0.15 \leq \frac{
 {\rm BR}(B_u\rightarrow\tau \nu_{\tau})_{\rm MSSM}}
 {{\rm BR}(B_u\rightarrow \tau \nu_{\tau})_{\rm SM}}
        \leq 2.41 \; (3\sigma)~~&\cite{Asner:2010qj}
\label{constraints}        
\end{eqnarray} 
 
In regard of the muon anomalous magnetic moment we require the solutions are at least as consistent as the Standard Model prediction \cite{Davier:2010nc}. 

In addition to those mentioned above, another constraint implied from the dark matter (DM) observations significantly limits the parameter space. It requires the lightest supersymmetric particle (LSP) stable and of no electric or color charge, which excludes the regions leading to $\tilde{\tau}$ or $\tilde{t}$ LSP solutions. On the other hand, even if a solution does not satisfy the DM observations, it can still survive in conjunction with other form(s) of DM \cite{Baer:2012by}. Therefore, we do not impose the DM constraints in our scan and we do not require the solutions to yield neutralino LSP. 
 
\section{Mass Spectrum in BLSSM Parameter Space}
\label{sec:results}

\begin{figure}[ht!]
\includegraphics[scale=0.45]{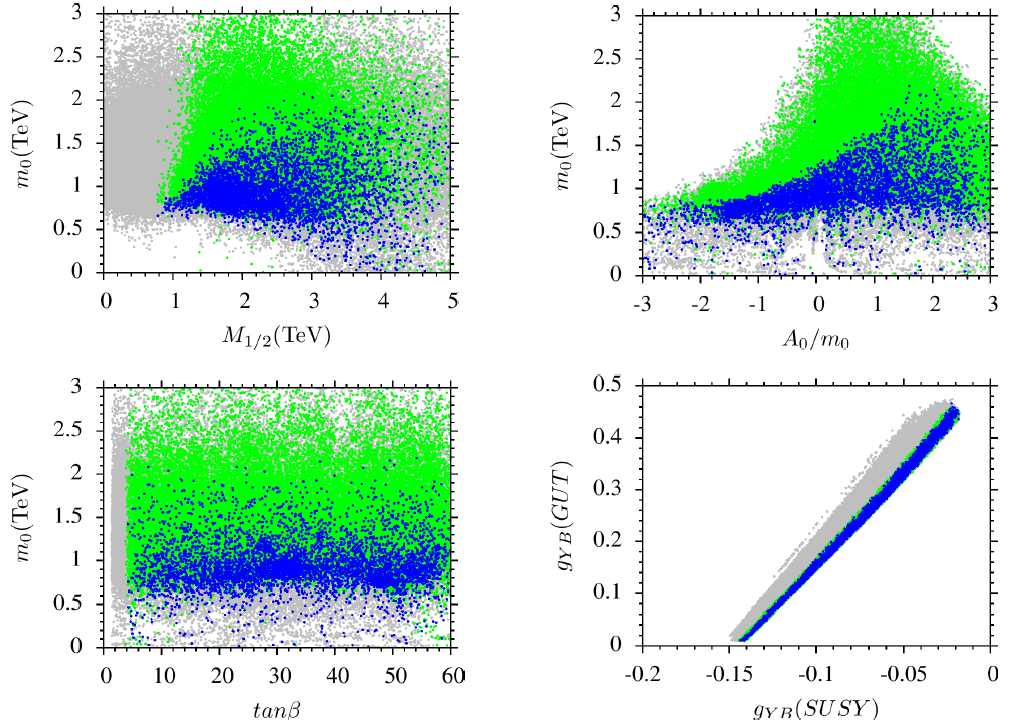}
\caption{Plots in $m_{0}-M_{1/2}$, $m_{0}-A_{0}/m_{0}$, $m_{0}-\tan\beta$, and $g_{YB}({\rm GUT})-g_{YB}({\rm SUSY})$ planes. All points are consistent with REWSB. Green points satisfy the mass bounds and the constraints from the rare B-decays. Blue points form a subset of green, and they represent solutions with $m_{h_{2}} \leq 150$ GeV.}
\label{fig2}
\end{figure}

\begin{figure}[ht!]
\includegraphics[scale=0.45]{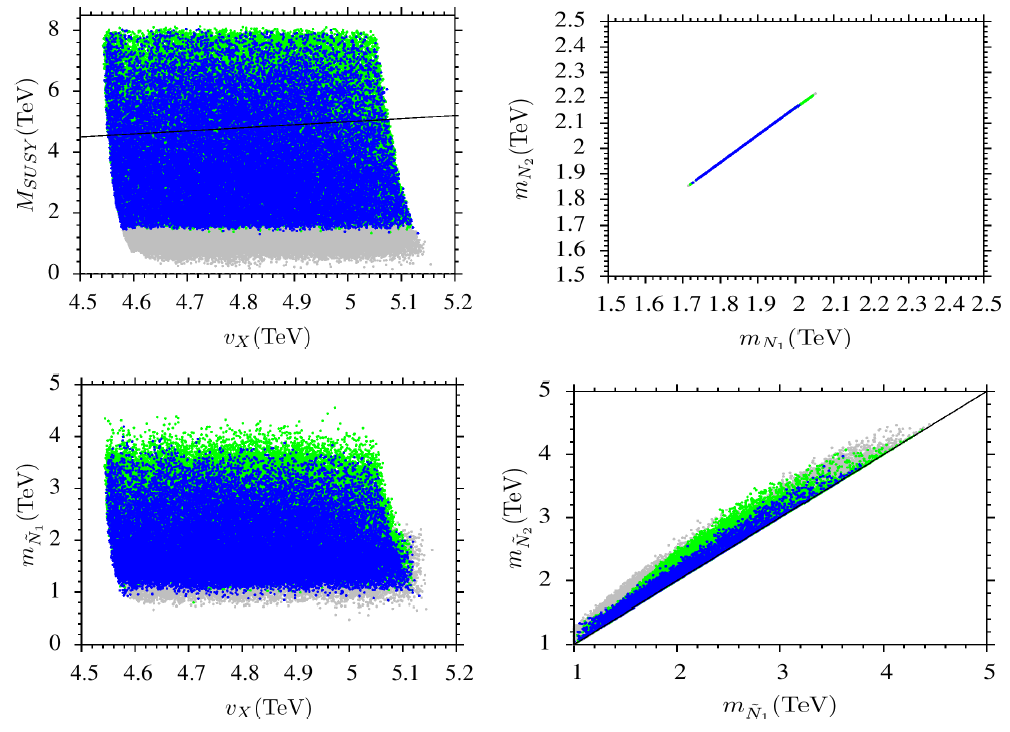}
\caption{Plots in the $M_{{\rm SUSY}}-v_{X}$, $m_{N_{2}}-m_{N_{1}}$, $M_{\tilde{N}_{1}}-v_{X}$, and $m_{\tilde{N_{2}}}-m_{\tilde{N_{1}}}$ planes. The color coding is the same as Figure \ref{fig2}. The solid line in the $M_{{\rm SUSY}}-v_{X}$ plane indicates the regions where $M_{{\rm SUSY}}=v_{X}$.}
\label{lepto1}
\end{figure}

\begin{figure}[ht!]
\includegraphics[scale=0.45]{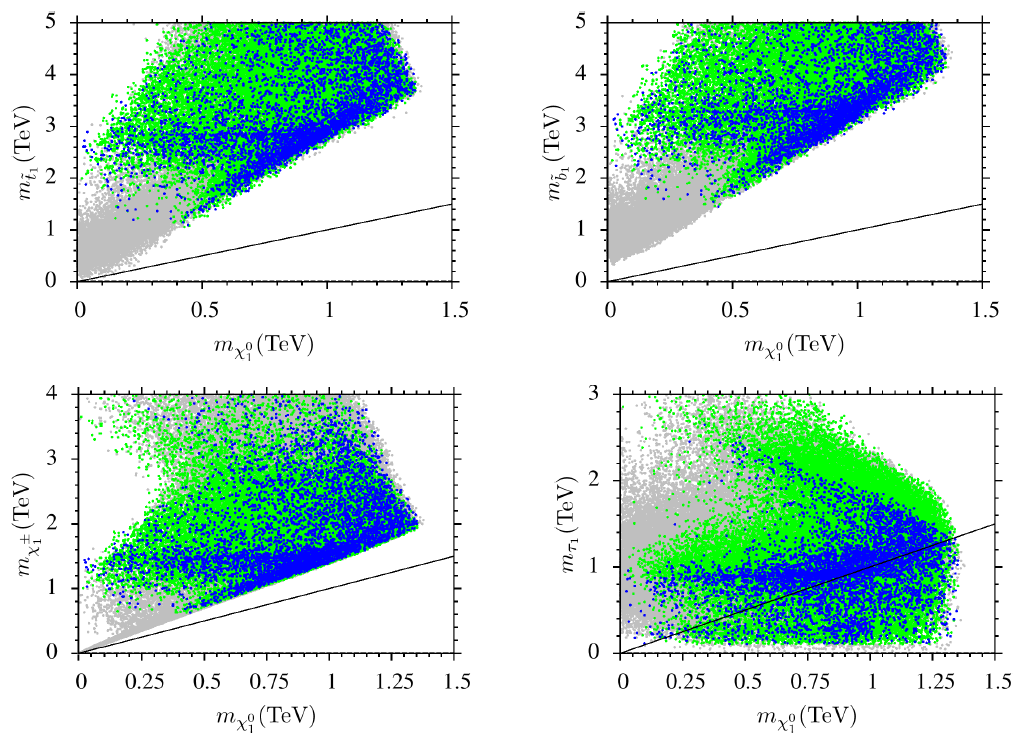}
\caption{Plots in $m_{\tilde{t}_{1}}-m_{\tilde{\chi}_{1}^{0}}$, $m_{\tilde{b}_{1}}-m_{\tilde{\chi}_{1}^{0}}$, $m_{\tilde{\tau}_{1}}-m_{\tilde{\chi}_{1}^{0}}$, and $m_{\tilde{\chi}^{\pm}_{1}}-m_{\tilde{\chi}_{1}^{0}}$ planes. The color coding is the same as Figure \ref{fig2}. In addition, the solid line shows the degenerate mass region in each plane.}
\label{fig3}
\end{figure}

\begin{figure}[ht!]
\includegraphics[scale=0.45]{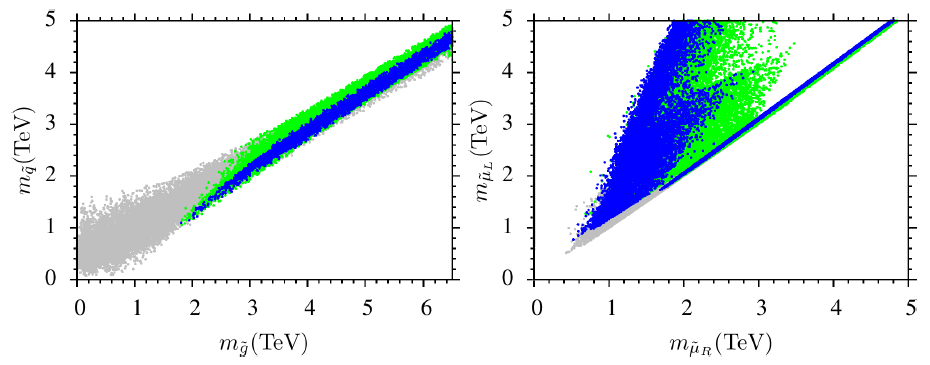}
\caption{Plots in $m_{\tilde{q}}-m_{\tilde{g}}$ and $m_{\tilde{\mu}_{L}}-m_{\tilde{\mu}_{R}}$ planes. The color coding is the same as Figure \ref{fig2}.}
\label{fig4}
\end{figure}

In this section, we present the results for the mass spectrum obtained from the scan over the parameter space given in Eq.(\ref{paramSP}). Figure \ref{fig2} displays the regions with plots in $m_{0}-M_{1/2}$, $m_{0}-A_{0}/m_{0}$, $m_{0}-\tan\beta$, and $g_{YB}({\rm GUT})-g_{YB}({\rm SUSY})$ planes. All points are consistent with REWSB. Green points satisfy the mass bounds and the constraints from the rare B-decays. Blue points form a subset of green, and they represent solutions with $m_{h_{2}} \leq 150$ GeV. As seen from the $m_{0}-M_{1/2}$ plane, the condition for the second Higgs boson lighter than 150 GeV (blue) excludes significant portion of the LHC allowed region (green). For $M_{1/2} \sim 1$ TeV, $m_{0}$ is restricted to a narrow range at about 500 GeV, and this range opens up to 2 TeV for heavier gaugino masses. This interplay can be partially understood with the heavy gaugino effect on the singlet Higgs mass. Even though it has very light masses at the GUT scale, the singlet Higgs boson mass is raised by the heavy $M_{B-L}$ such that $m_{h_{2}} \lesssim 150$ GeV. On the other hand, for the large values of $m_{0}$, which means heavy $m_{\singone}$ and $m_{N}$, as seen from Eq.(\ref{RGEsingone}), these masses reduce the singlet Higgs boson mass. The results in the $m_{0}-M_{1/2}$ plane shows that $m_{0}$ reaches its highest values when $m_{0}\approx M_{1/2}\sim 2$ TeV. On the other hand, the $m_{0}-A_{0}/m_{0}$ panel shows that the heavy gaugino mass cannot explain the results fully, the regions with larger $m_{0}$ values requires positive SSB trilinear scalar interaction term and when $A_{0}/m_{0} \gtrsim 1.5$ , $m_{0}$ can be as large as 2 TeV and the solutions can still yield two Higgs boson with mass $\leq 150$ GeV. When $A_{0}$ is negative, the RGE evolution of $A_{N}$ has an increasing slope, and its contribution to the singlet Higgs boson takes over the heavy gaugino effect. Therefore the solutions with large $A_{N}$ needs to be restricted with the low $m_{0}$ and $M_{1/2}$ values. The $m_{0}-\tan\beta$ plane shows that it is possible to find solutions with $m_{h_{2}}\leq 150$ GeV for almost all values of $\tan\beta$. Finally the $g_{YB}({\rm GUT})-g_{YB}({\rm SUSY})$ plane represents our results in regard of the gauge kinetic mixing. Even though we vary it in the perturbative level at the GUT scale, its low scale value is found in the range ($-0.15-0$)  

In Figure \ref{lepto1} we present our results in the  $M_{{\rm SUSY}}-v_{X}$, $m_{N_{2}}-m_{N_{1}}$, $M_{\tilde{N}_{1}}-v_{X}$, and $m_{\tilde{N_{2}}}-m_{\tilde{N_{1}}}$ planes. The color coding is the same as Figure \ref{fig2}. The solid line in the $M_{{\rm SUSY}}-v_{X}$ plane indicates the regions where $M_{{\rm SUSY}}=v_{X}$. According to our results, the breaking of $U(1)_{B-L}$ happens at about $v_{X}\approx 5$ TeV. Since $U(1)_{B-L}$ is no more the symmetry in the model, and the existence of the right-handed neutrinos can trigger baryon and lepton number violating processes, which can be considered as a source for the baryon asymmetry in the Universe. Assuming that the supsersymmetric particles all decouple below $M_{{\rm SUSY}}$, the $M_{{\rm SUSY}}-v_{X}$ plane shows that $U(1)_{B-L}$ symmetry breaking can be realized in both supersymmetric regime ($v_{X} > M_{{\rm SUSY}}$) and non-supersymmetric regime ($v_{X} < M_{{\rm SUSY}}$).  In the non-supersymmetric regime, the baryon and lepton violating processes rely on the right-handed neutrinos. Since the Yukawa coupling associated with the neutrinos is very small ($Y_{\nu} \sim 10^{-7}$), the thermal leptogenesis can provide sufficient baryon assymmetry when the right-handed neutrinos are degenerate in mass \cite{Abbas:2007ag,Flanz:1994yx}. As shown in the $m_{N_{2}}-m_{N_{1}}$ plane, the right-handed neutrino masses ($\sim 1.7-2.2$ TeV) are nearly degenerate . In addition to the right-handed neutrinos, the sneutrino-antisneutrino can be counted as another source in the supersymmetric regime \cite{Grossman:2003jv}. After the right-handed neutrinos decouple, $B-L$ symmetry is restored globally. 

Figure \ref{fig3} represents the results for the sparticle mass spectrum with plots in $m_{\tilde{t}_{1}}-m_{\tilde{\chi}_{1}^{0}}$, $m_{\tilde{b}_{1}}-m_{\tilde{\chi}_{1}^{0}}$, $m_{\tilde{\chi}^{\pm}_{1}}-m_{\tilde{\chi}_{1}^{0}}$ and $m_{\tilde{\tau}_{1}}-m_{\tilde{\chi}_{1}^{0}}$ planes. The color coding is the same as Figure \ref{fig2}. In addition, the solid line shows the degenerate mass region in each plane. As is seen from the $m_{\tilde{t}_{1}}-m_{\tilde{\chi}_{1}^{0}}$ and $m_{\tilde{b}_{1}}-m_{\tilde{\chi}_{1}^{0}}$ planes, $m_{\tilde{t}_{1}} \gtrsim 1$ and $m_{\tilde{b}_{1}}\gtrsim 1.5$ TeV, and these masses are mostly required to realize the SM-like Higgs boson mass at about 125 GeV. Moreover, the $m_{\tilde{\chi}^{\pm}_{1}}-m_{\tilde{\chi}_{1}^{0}}$ plane shows that the lightest chargino cannot be lighter than 600 GeV. Even though we do not require the neutralino to be LSP, it is found much lighter than other sparticles except stau. The $m_{\tilde{\tau}_{1}}-m_{\tilde{\chi}_{1}^{0}}$ plane represents the stau mass along with the neutralino mass, and it can be lighter than neutralino as well as being much heavier. One can constrain the stau mass further by the prompt decay of stau to gravitino in the case of gravitino LSP \cite{Calibbi:2014pza}.

We continue with Figure \ref{fig4} to present our results for the sparticle spectrum with plots in $m_{\tilde{q}}-m_{\tilde{g}}$ and $m_{\tilde{\mu}_{L}}-m_{\tilde{\mu}_{R}}$ planes. The color coding is the same as Figure \ref{fig2}. The $m_{\tilde{q}}-m_{\tilde{g}}$ shows that the squarks from the first two families and gluino should be heavier than 2 TeV. Even though we impose a mass bound on gluino at about 1.8 TeV, the other LHC results mentioned in Section \ref{sec:scan} constrain gluino mass further to about 2 TeV (green). Imposing the condition that $m_{h_{2}} \leq 150$ GeV (blue) does not constrain the gluino or squark masses strictly. Similarly the results for the smuon masses are represented in the $m_{\tilde{\mu}_{L}}-m_{\tilde{\mu}_{R}}$ plane. According to the our results, the lightest left- and right-handed smuon masses are about 1 TeV. In this case, one can expect relatively better result for the muon anomalous magnetic moment (muon $g-2$), but since the supersymmetric contributions are more or less suppressed by the smuon masses, the results for the muon $g-2$ hardly reach to $2\sigma$ band of the experimental results.  

\begin{figure}[t!]
\includegraphics[scale=0.45]{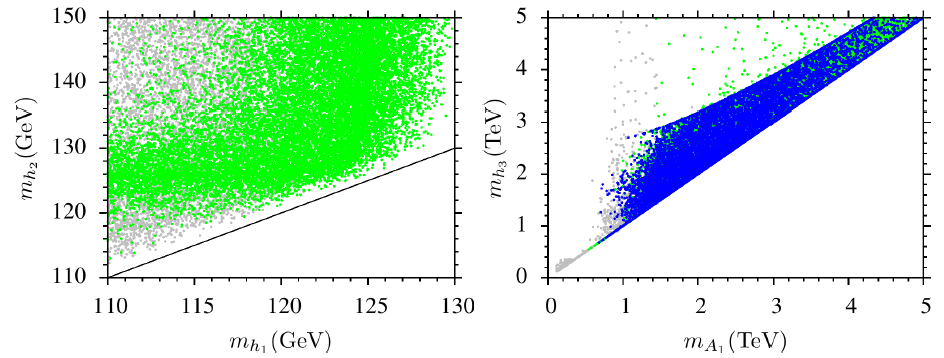}
\caption{Plots in $m_{h_{2}}-m_{h_{1}}$ and $m_{h_{3}}-m_{A_{1}}$ planes. The color coding is the same as Figure \ref{fig2} except that the Higgs mass bound in green is not applied in the $m_{h_{2}}-m_{h_{1}}$ plane since $m_{h_{1}}$ is plotted in one axis. The diagonal line represents the mass degeneracy.}
\label{fig5}
\end{figure}

Finally we display our results for the mass spectrum of the Higgs bosons in Figure \ref{fig5} with plots in $m_{h_{2}}-m_{h_{1}}$ and $m_{h_{3}}-m_{A_{1}}$ planes. The color coding is the same as Figure \ref{fig2} except that the Higgs mass bound in green is not applied in the $m_{h_{2}}-m_{h_{1}}$ since $m_{h_{1}}$ is plotted in one axis. The diagonal line represents the mass degeneracy. The $m_{h_{2}}-m_{h_{1}}$ plane shows that there are plenty of solutions with $m_{h_{1}},~m_{h_{2}} \leq 150$ GeV. Moreover, following the diagonal line we can see that it is also possible to find the lightest two Higgs boson with almost degenerate at about $m_{h_{1}}\approx m_{h_{2}} \sim 125$ GeV. The other Higgs bosons are found rather heavy ($\gtrsim 1$ TeV) as shown in the $m_{h_{3}}-m_{A_{1}}$ plane. 

\section{Higgs Decays}
\label{subsec41}

We have represented the mass spectrum in BLSSM in the previous section. As mentioned, BLSSM provide an extra Higgs boson which can be lighter than 150 GeV, and even two Higgs bosons can be degenerate at about 125 GeV. With the mixing between two Higgs fields this region can provide a relatively rich phenomenology for the Higgs decays. In this section, we present our results for the Higgs decays into two photons and four leptons.

\subsection{$h\rightarrow \gamma \gamma$}
\label{subsubsec411}

\begin{figure}[ht!]
\includegraphics[scale=0.45]{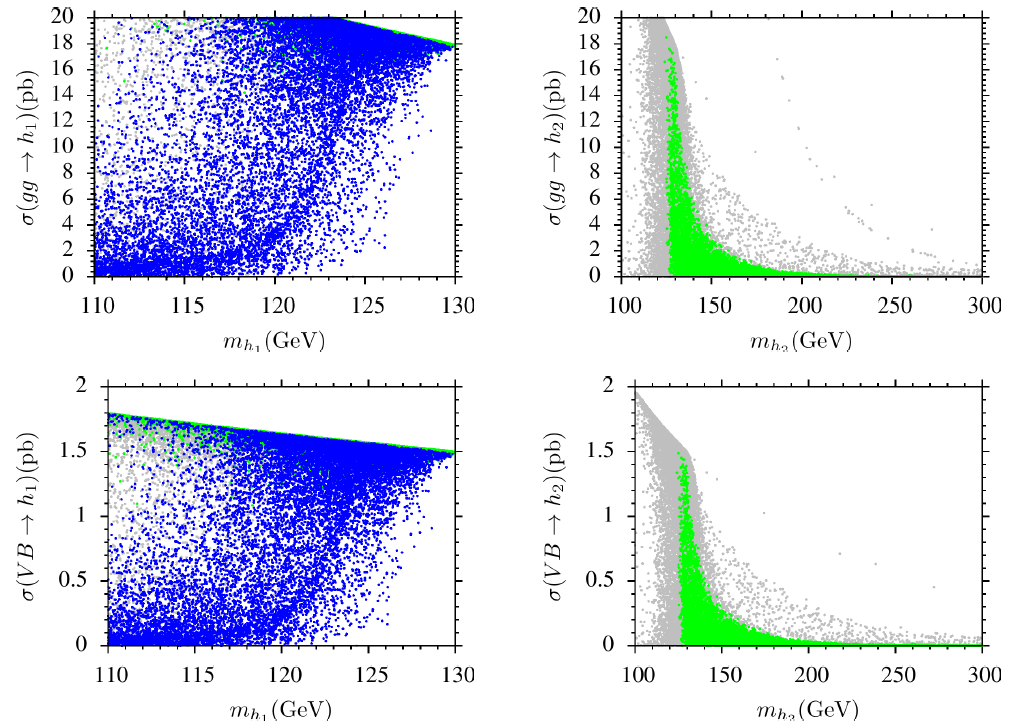}
\caption{Plots for the Higgs boson production cross-section through GGF (top panel) and VBF (bottom panel) in the $\sigma(gg\rightarrow h_{1})-m_{h_{1}}$, $\sigma(VB\rightarrow h_{1})-m_{h_{1}}$, $\sigma(gg\rightarrow h_{2})-m_{h_{2}}$ and $\sigma(VB\rightarrow h_{2})-m_{h_{2}}$ planes. The color coding is the same as Figure \ref{fig2}, except we do not apply the SM Higgs boson constraint ($m_{h_{1}} \sim 125$ GeV) to the left panel, since $m_{h_{1}}$ is directly plotted here. Similarly, the condition $m_{h_{2}}\leq 150$ GeV, represented by the blue region, is not applied to the right panels, since $m_{h_{2}}$ is on the horizontal axis.}
\label{fig6}
\end{figure}

\begin{figure}[ht!]
\includegraphics[scale=0.45]{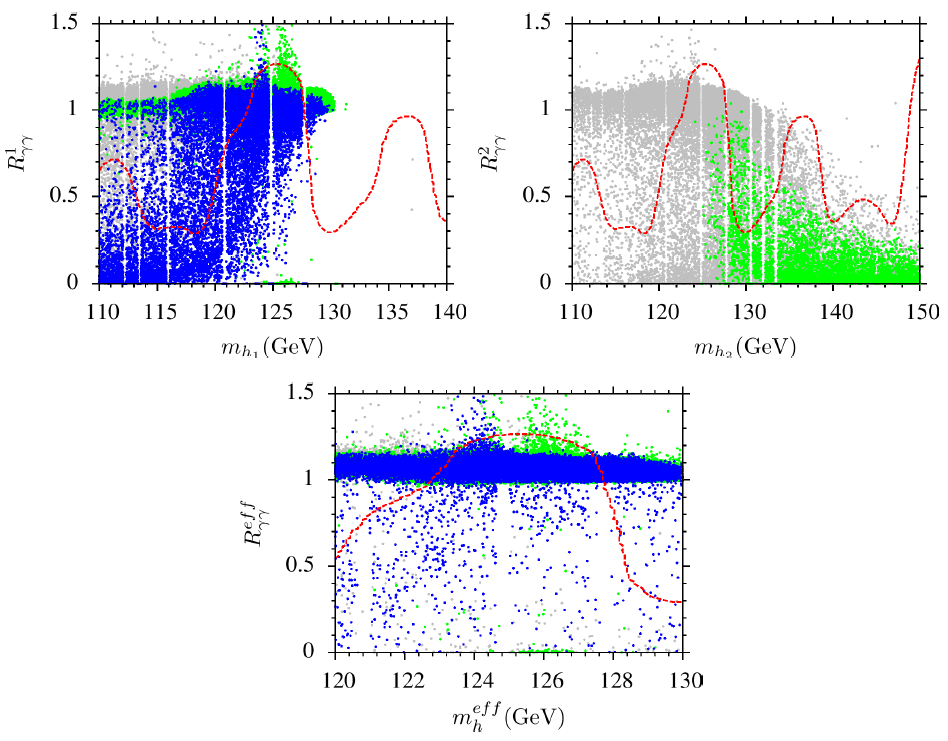}
\caption{Plots in $R_{\gamma\gamma}^{1}-m_{h_{1}}$, $R_{\gamma\gamma}^{2}-m_{h_{2}}$ and $R_{\gamma \gamma}^{{\rm eff}}-m_{h}^{{\rm eff}}$ planes. The color coding is the same as Figure \ref{fig6}. The red dashed line indicates the observed cross-section in $h\rightarrow \gamma \gamma$ normalized to the SM prediction.} 
\label{fig7}
\end{figure}

The sparticles shown in Figure \ref{fig3} contribute to the loop induced coupling between the Higgs boson and two photons in SUSY models. Since their contributions are inversely proportional to their masses, the contributions from stop and sbottom are suppressed by their heavy masses. The main contribution comes from the stau, since its mass can be as low as 100 GeV. In addition, chargino contribution can be counted as a correction. Moreover, since the second Higgs boson mass lighter than 150 GeV can be realized, there is also an induced coupling between $h_{2}$ and two photons.  One can quantify the excess relative to the SM prediction in $h\rightarrow \gamma \gamma$ with the parameter $R_{\gamma\gamma}^{i}$ defined as 

\begin{equation}
 R_{\gamma \gamma}^{i}=\frac{\sigma(pp\rightarrow h_{i})\times {\rm BR}(h_{i}\rightarrow \gamma \gamma)}{\sigma(pp\rightarrow h)_{{\rm SM}}\times {\rm BR}(h\rightarrow \gamma\gamma)_{{\rm SM}}}
 \label{ecxess1}
\end{equation} 
where $\sigma(pp\rightarrow h_{i})$ denotes the production cross-section of the Higgs boson $h_{i}$, and ${\rm BR}(h_{i}\rightarrow \gamma \gamma)$ is the branching ratio of the process in which the Higgs boson decays into two photons. The definitions for the terms in the denominator are the same, but they represent the SM predictions for the same process. 

Eq.(\ref{ecxess1}) reveals the importance of the Higgs boson production at the LHC as well as the loop induced coupling between the Higgs bosons and photons. Since the Higgs boson couplings to the matter fields in the first two families are negligible, the main contributions to $\sigma(pp\rightarrow h_{i})$ come from gluon fusion (GGF), vector boson fusion (VBF), associated vector boson-Higgs (VH) production and higgs production along with the top quark pair (ttH). Figure \ref{fig6} displays plots for the Higgs boson production cross-section through GGF (top panel) and VBF (bottom panel) in the $\sigma(gg\rightarrow h_{1})-m_{h_{1}}$, $\sigma(VB\rightarrow h_{1})-m_{h_{1}}$, $\sigma(gg\rightarrow h_{2})-m_{h_{2}}$ and $\sigma(VB\rightarrow h_{2})-m_{h_{2}}$ planes. The color coding is the same as Figure \ref{fig2}, except we do not apply the SM Higgs boson constraint ($m_{h_{1}} \sim 125$ GeV) to the left panel, since $m_{h_{1}}$ is directly plotted here. Similarly,the condition $m_{h_{2}}\leq 150$ GeV, represented by the blue region, is not applied to the right panels, since $m_{h_{2}}$ is on the horizontal axis. As seen from the plots of Figure \ref{fig6}, GGF dominates in the Higgs boson production at the LHC as happened for the SM Higgs boson. However, while GGF yields a production cros-section of the order about $10^{2}$ pb in the SM \cite{Djouadi:2005gi} and MSSM \cite{Djouadi:2005gj}, in BLSSM the GGF cross-section is found at about 20 pb at most. This is because the Higgs boson couplings are diminished by $\sin\alpha$ and $\cos\alpha$, where $\alpha$ measures the mixing between the Higgs fields. As shown in the $\sigma(gg\rightarrow h_{1})-m_{h_{1}}$ plane, $h_{1}$ behaves mostly like the SM Higgs boson, while $h_{2}$ can share this behavior when $m_{h_{2}} \lesssim 150$ GeV. As seen from the $\sigma(gg\rightarrow h_{2})-m_{h_{2}}$ plane, the $h_{2}$ production has a sharp fall for relatively heavier mass scales, and finally it drops to zero for $m_{h_{2}} \gtrsim 200$ GeV. It is because the second lightest higgs boson is mostly formed by the BLSSM Higgs fields, which are SM-singlets, as the mass difference between the two lightest Higgs bosons increases.  A similar discussion can hold for the VBF as shown in the bottom plane of Figure \ref{fig6}. VBF is usually the production channel with the second larger contribution, and it is one order of magnitude smaller than the GGF results.

We present our results for the possible excesses in $h_{i}\rightarrow \gamma \gamma$ with plots in $R_{\gamma\gamma}^{1}-m_{h_{1}}$, $R_{\gamma\gamma}^{2}-m_{h_{2}}$ and $R_{\gamma \gamma}^{{\rm eff}}-m_{h}^{{\rm eff}}$ planes. The color coding is the same as Figure \ref{fig6}. The red dashed line indicates the observed cross-section in $h\rightarrow \gamma \gamma$ normalized to the SM prediction. As seen from the $R_{\gamma\gamma}^{1}-m_{h_{1}}$ plane, BLSSM yields plenty of solutions which can feed the excess in $h\rightarrow \gamma \gamma$ for both $m_{h_{2}} \leq 150$ GeV (blue) and $m_{h_{2}} \geq 150$ GeV (green). These solutions can be explained by effects of the light staus and relatively light charginos as shown in Figure \ref{fig3}. In addition to the light sparticles, also the second lightest Higgs boson mass can be realized as nearly degenerate with $m_{h_{1}}\approx 125$ GeV, and it can be seen from the $R_{\gamma\gamma}^{2}-m_{h_{2}}$ plane that it can provide some cross-section in $h\rightarrow \gamma \gamma$ as much as the SM ($R^{2}_{\gamma\gamma}\sim 1$). In this region, we have two Higgs bosons of mass about 125 GeV, and both contribute to the cross-section of $h\rightarrow \gamma \gamma$. If we define $m_{h}^{{\rm eff}}$ and $R_{\gamma \gamma}^{{\rm eff}}$ as

\begin{equation}
m_{h}^{{\rm eff}}=\frac{m_{h_{1}}R_{\gamma\gamma}^{1}+ m_{h_{2}}R_{\gamma\gamma}^{2}}{R_{\gamma \gamma}^{1}+R_{\gamma \gamma}^{2}},\hspace{0.5cm} R_{\gamma \gamma}^{{\rm eff}}=R_{\gamma \gamma}^{1}+R_{\gamma \gamma}^{2}
\end{equation}
the predicted effective cross-section by many solutions are lifted up to region where $R_{\gamma \gamma}^{{\rm eff}} \gtrsim 1$ for $m_{h}^{{\rm eff}} \sim 125$ GeV, as seen from the $R_{\gamma \gamma}^{{\rm eff}}-m_{h}^{{\rm eff}}$ plane.

Before concluding it should be noted that the second lightest higgs boson can be accounted for the other peaks at about 137 GeV and 145 GeV observed in the experiments. As seen from the $R_{\gamma\gamma}^{2}-m_{h_{2}}$ panel, the solutions may relatively provide some non-zero cross-sections at these mass scales. However, the solutions around the second peak at 137 GeV are excluded by the Higgs boson constraint. Since we have restricted ourselves with the universal boundary conditions at $M_{{\rm GUT}}$, these predictions can be ameliorated by imposing non-universality.

\subsection{$h\rightarrow 4l$}
\label{subsubsec42}

\begin{figure}[ht!]
\includegraphics[scale=0.45]{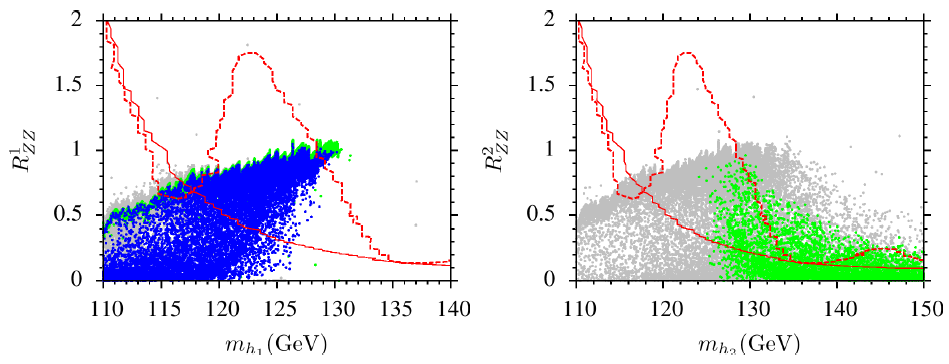}
\caption{Plots in $R_{ZZ}^{1}-m_{h_{1}}$ and $R_{ZZ}^{2}-m_{h_{2}}$. The color coding is the same as Figure \ref{fig6}. The dashed line indicates the observed cross-section, while the solid line represents the expected cross-section without the Higgs boson.}
\label{fig8}
\end{figure}

A similar discussion can be followed for the process in which the Higgs boson decays into four leptons. In the SM, this process is mediated via two Z-bosons, each of which eventually decays into a lepton pair. In BLSSM, such decays can include also $Z'$, but due to its heavy mass ($m_{Z'}=2.5$ TeV in our work), such processes are highly suppressed. Hence, the difference in $h\rightarrow 4l$ between BLSSM and the observation basically come from the Higgs boson decays into two $Z-$bosons. Figure \ref{fig8} represents our results with plots in  $R_{ZZ}^{1}-m_{h_{1}}$ and $R_{ZZ}^{2}-m_{h_{2}}$. The color coding is the same as Figure \ref{fig6}. The dashed line indicates the observed cross-section, while the solid line represents the expected cross-section without the Higgs boson. In contrast to the Higgs decays into two photons, BLSSM's predictions can be only as good as ones in the SM, even in the case of the degenerate Higgs bosons. On the other hand, if one considers the second peak observed at $m_{h}\sim 145$ GeV, it can be seen from the $R_{ZZ}^{2}-m_{h_{2}}$ plane, the second Higgs boson can nicely fill the region around this peak. 

\section{Conclusion}
\label{sec:conc}

We presented the predictions on the mass spectrum and Higgs boson decays in the 
BLSSM framework with universal boundary conditions. We briefly mentioned about the right-handed neutrino sector. The radiative breaking of $U(1)_{B-L}$ symmetry happens at about 5 TeV below which $B-L$ is no more the conserved symmetry and the right-handed neutrinos can trigger baryon and lepton number violating process till they decouple from the SM sector at $1.7-2.2$ TeV. Radiative breaking of $B-L$ symmetry can happen in both supersymemtric ($v_{X} > M_{{\rm SUSY}}$) and non-supersymmetric ($v_{X} < M_{{\rm SUSY}}$). The sneutrino-antisneutrino mixing can be counted as another source for baryon and lepton asymmetry in the Universe.

We found the stop and sbottom masses heavier than 1.5 TeV, and gluino mass greater than 2 TeV. The color sector is required to be heavy in order to realize the SM-like Higgs boson consistent with the observations. Even though BLSSM's predictions for the Higgs boson are similar to MSSM, it predicts another Higgs boson, which can be lighter than 150 GeV, and even degenerate with the lightest CP-even Higgs boson at about 125 GeV. Besides light staus ($\gtrsim 100$ GeV), the second Higgs boson also contributes to the Higgs decay processes in the presence of gauge kinetic mixing. We showed that the excess in $h\rightarrow \gamma \gamma$ at about 125 GeV mass scale can be realized. The solutions which can provide an excess at 137 GeV and 145 GeV in this process are rather excluded by the 125 GeV Higgs boson constraint. Such solutions can be cured by considering non-universal boundary conditions in BLSSM. In addition, we concluded that the BLSSM predictions for $h\rightarrow 4l$ are only as good as the SM, but it is eligible to fit the second excess at about 145 GeV.

\vspace{0.2cm}
\textbf{Acknowledgement}
We would like to thank Durmus Ali Demir and Shaaban Khalil for useful discussions and comments. This work is supported by The Scientific and Technological Research Council of Turkey (TUBITAK) Grant no. MFAG-114F461. CSU acknowledges the support H2020-MSCA-RISE-2014 Grant no. 645722 (NonMinimalHiggs), with which some parts of this work were completed at the University of Southampton. This work used Extreme Science and Engineering Discovery Environment (XSEDE), which is supported by the National Science Foundation grant number OCI-1053575. Part of numerical calculations reported in this paper were performed at the National Academic Network and Information Center (ULAKBIM) of TUBITAK, High Performance and Grid Computing Center (TRUBA resources).


\begin{thebibliography}{99}

%\cite{:2012gk}
\bibitem{:2012gk}
  G.~Aad {\it et al.}  [ATLAS Collaboration],
  %``Observation of a new particle in the search for the Standard Model Higgs boson with the ATLAS detector at the LHC,''
  Phys.\ Lett.\ B {\bf 716}, 1 (2012).
 % [arXiv:1207.7214 [hep-ex]].
  %%CITATION = ARXIV:1207.7214;%%

%\cite{:2012gu}
\bibitem{:2012gu}
  S.~Chatrchyan {\it et al.}  [CMS Collaboration],
  %``Observation of a new boson at a mass of 125 GeV with the CMS experiment at the LHC,''
  Phys.\ Lett.\ B {\bf 716}, 30 (2012).
 % [arXiv:1207.7235 [hep-ex]].
  %%CITATION = ARXIV:1207.7235;%%

  %\cite{Gildener:1976ai}
\bibitem{Gildener:1976ai}
  E.~Gildener,
  %``Gauge Symmetry Hierarchies,''
  Phys.\ Rev.\ D {\bf 14}, 1667 (1976);
  %%CITATION = PHRVA,D14,1667;%%
  %499 citations counted in INSPIRE as of 12 Sep 2014
  %\cite{Gildener:1979dd}
%\bibitem{Gildener:1979dd}
  E.~Gildener,
  %``Gauge Symmetry Hierarchies Revisited,''
  Phys.\ Lett.\ B {\bf 92}, 111 (1980);
  %%CITATION = PHLTA,B92,111;%%
  %54 citations counted in INSPIRE as of 12 Sep 2014
  %\cite{Veltman:1980mj}
  S.~Weinberg,
  %``Gauge Hierarchies,''
  Phys.\ Lett.\ B {\bf 82}, 387 (1979);
  %%CITATION = PHLTA,B82,387;%%
  %230 citations counted in INSPIRE as of 04 Sep 2013
L.~Susskind,
  %``Dynamics of Spontaneous Symmetry Breaking in the Weinberg-Salam Theory,''
  Phys.\ Rev.\ D {\bf 20}, 2619 (1979);
  %%CITATION = PHRVA,D20,2619;%%
  %1924 citations counted in INSPIRE as of 04 Sep 2013
%\bibitem{Veltman:1980mj}
  M.~J.~G.~Veltman,
  %``The Infrared - Ultraviolet Connection,''
  Acta Phys.\ Polon.\ B {\bf 12}, 437 (1981);
  %%CITATION = APPOA,B12,437;%%
  %517 citations counted in INSPIRE as of 12 Sep 2014
\bibitem{hinstability}
G.~Degrassi, S.~Di Vita, J.~Elias-Miro, J.~R.~Espinosa, G.~F.~Giudice, G.~Isidori and A.~Strumia,
  %``Higgs mass and vacuum stability in the Standard Model at NNLO,''
  JHEP {\bf 1208}, 098 (2012)
  [arXiv:1205.6497 [hep-ph]];
  F.~Bezrukov, M.~Y.~.Kalmykov, B.~A.~Kniehl and M.~Shaposhnikov,
  %``Higgs Boson Mass and New Physics,''
  JHEP {\bf 1210}, 140 (2012)
  [arXiv:1205.2893 [hep-ph]];
  %%CITATION = ARXIV:1205.2893;%%
  %137 citations counted in INSPIRE as of 18 May 2014
  %%CITATION = ARXIV:1205.6497;%%
  %263 citations counted in INSPIRE as of 18 May 2014
D.~Buttazzo, G.~Degrassi, P.~P.~Giardino, G.~F.~Giudice, F.~Sala, A.~Salvio and A.~Strumia,
  %``Investigating the near-criticality of the Higgs boson,''
  JHEP {\bf 1312}, 089 (2013)
  [arXiv:1307.3536].
  %%CITATION = ARXIV:1307.3536;%%
  %105 citations counted in INSPIRE as of 18 May 2014

%\cite{Khachatryan:2015cwa}
\bibitem{Khachatryan:2015cwa}
  V.~Khachatryan {\it et al.}  [CMS Collaboration],
  %``Search for a Higgs boson in the mass range from 145 to 1000 GeV decaying to a pair of W or Z bosons,''
  arXiv:1504.00936 [hep-ex].
  %%CITATION = ARXIV:1504.00936;%%
  %7 citations counted in INSPIRE as of 27 May 2015

  %\cite{CMS:ril}
\bibitem{CMS:ril}
  [CMS Collaboration],
  %``Updated measurements of the Higgs boson at 125 GeV in the two photon decay channel,''
  CMS-PAS-HIG-13-001.
  %%CITATION = CMS-PAS-HIG-13-001;%%
  %233 citations counted in INSPIRE as of 27 May 2015

%\cite{Chatrchyan:2013mxa}
\bibitem{Chatrchyan:2013mxa}
  S.~Chatrchyan {\it et al.}  [CMS Collaboration],
  %``Measurement of the properties of a Higgs boson in the four-lepton final state,''
  Phys.\ Rev.\ D {\bf 89}, no. 9, 092007 (2014)
  [arXiv:1312.5353 [hep-ex]].
  %%CITATION = ARXIV:1312.5353;%%
  %233 citations counted in INSPIRE as of 27 May 2015

  %\cite{Ferreira:2012nv}
\bibitem{Ferreira:2012nv}
For an incomplete list, see

  P.~M.~Ferreira, R.~Santos, H.~E.~Haber and J.~P.~Silva,
  %``Mass-degenerate Higgs bosons at 125 GeV in the two-Higgs-doublet model,''
  Phys.\ Rev.\ D {\bf 87}, 055009 (2013)
  [arXiv:1211.3131 [hep-ph]];
  %%CITATION = ARXIV:1211.3131;%%
  %54 citations counted in INSPIRE as of 04 juin 2015
%\cite{Han:2013mga}
%\bibitem{Han:2013mga}
  T.~Han, T.~Li, S.~Su and L.~T.~Wang,
  %``Non-Decoupling MSSM Higgs Sector and Light Superpartners,''
  JHEP {\bf 1311}, 053 (2013)
  [arXiv:1306.3229 [hep-ph]];
  %%CITATION = ARXIV:1306.3229;%%
  %16 citations counted in INSPIRE as of 04 juin 2015
%\cite{Ke:2012zq}
%\bibitem{Ke:2012zq}
  J.~Ke, H.~Luo, M.~x.~Luo, K.~Wang, L.~Wang and G.~Zhu,
  %``Revisit to Non-decoupling MSSM,''
  Phys.\ Lett.\ B {\bf 723}, 113 (2013)
  [arXiv:1211.2427 [hep-ph]].
  %%CITATION = ARXIV:1211.2427;%%
  %18 citations counted in INSPIRE as of 04 juin 2015

%\cite{Carena:2012mw}
\bibitem{Carena:2012mw}
  M.~Carena, S.~Gori, I.~Low, N.~R.~Shah and C.~E.~M.~Wagner,
  %``Vacuum Stability and Higgs Diphoton Decays in the MSSM,''
  JHEP {\bf 1302}, 114 (2013)
  [arXiv:1211.6136 [hep-ph]].
  %%CITATION = ARXIV:1211.6136;%%
  %59 citations counted in INSPIRE as of 12 Nov 2015

  %\cite{Demir:2014jqa}
\bibitem{Demir:2014jqa}
  D.~A.~Demir and C.~S.~Ün,
  %``Stop on Top: SUSY Parameter Regions, Fine-Tuning Constraints,''
  Phys.\ Rev.\ D {\bf 90}, 095015 (2014)
  [arXiv:1407.1481 [hep-ph]].
  %%CITATION = ARXIV:1407.1481;%%
  %1 citations counted in INSPIRE as of 12 Nov 2015

%\cite{Gogoladze:2013wva}
\bibitem{Gogoladze:2013wva}
  I.~Gogoladze, F.~Nasir and Q.~Shafi,
  %``SO(10) as a Framework for Natural Supersymmetry,''
  JHEP {\bf 1311}, 173 (2013)
  [arXiv:1306.5699 [hep-ph]].
  %%CITATION = ARXIV:1306.5699;%%
  %14 citations counted in INSPIRE as of 12 Nov 2015

%\cite{Aaij:2012nna}
\bibitem{Aaij:2012nna}
  R.~Aaij {\it et al.} [LHCb Collaboration],
  %``First Evidence for the Decay $B_s^0 \to \mu^+ \mu^-$,''
  Phys.\ Rev.\ Lett.\  {\bf 110}, no. 2, 021801 (2013)
  [arXiv:1211.2674 [hep-ex]].
  %%CITATION = ARXIV:1211.2674;%%
  %354 citations counted in INSPIRE as of 12 Nov 2015

%\cite{Bobeth:2013uxa}
\bibitem{Bobeth:2013uxa}
  C.~Bobeth, M.~Gorbahn, T.~Hermann, M.~Misiak, E.~Stamou and M.~Steinhauser,
  %``B_{s,d} -> l+ l- in the Standard Model with Reduced Theoretical Uncertainty,''
  Phys.\ Rev.\ Lett.\  {\bf 112}, 101801 (2014)
  [arXiv:1311.0903 [hep-ph]];
  %%CITATION = ARXIV:1311.0903;%%
  %114 citations counted in INSPIRE as of 12 Nov 2015
%\bibitem{Bobeth:2013tba}
  C.~Bobeth, M.~Gorbahn and E.~Stamou,
  %``Electroweak Corrections to $B_{s,d} \to \ell^+ \ell^-$,''
  Phys.\ Rev.\ D {\bf 89}, no. 3, 034023 (2014)
  [arXiv:1311.1348 [hep-ph]];
  %%CITATION = ARXIV:1311.1348;%%
  %38 citations counted in INSPIRE as of 12 Nov 2015
%\bibitem{Hermann:2013kca}
  T.~Hermann, M.~Misiak and M.~Steinhauser,
  %``Three-loop QCD corrections to $B_s \to \mu^+ \mu^-$,''
  JHEP {\bf 1312}, 097 (2013)
  [arXiv:1311.1347 [hep-ph]].
  %%CITATION = ARXIV:1311.1347;%%
  %40 citations counted in INSPIRE as of 12 Nov 2015

  %\cite{Gogoladze:2014vea}
\bibitem{Gogoladze:2014vea}
  I.~Gogoladze, B.~He, A.~Mustafayev, S.~Raza and Q.~Shafi,
  %``Effects of Neutrino Inverse Seesaw Mechanism on the Sparticle Spectrum in CMSSM and NUHM2,''
  JHEP {\bf 1405}, 078 (2014)
  [arXiv:1401.8251 [hep-ph]].
  %%CITATION = ARXIV:1401.8251;%%
  %2 citations counted in INSPIRE as of 12 Nov 2015

  %\cite{Langacker:2008yv}
\bibitem{Langacker:2008yv}
  P.~Langacker,
  %``The Physics of Heavy $Z^\prime$ Gauge Bosons,''
  Rev.\ Mod.\ Phys.\  {\bf 81}, 1199 (2009)
  [arXiv:0801.1345 [hep-ph]] and references therein.
  %%CITATION = ARXIV:0801.1345;%%
  %656 citations counted in INSPIRE as of 12 Nov 2015

%\cite{Wendell:2010md}
\bibitem{Wendell:2010md}
  R.~Wendell {\it et al.} [Super-Kamiokande Collaboration],
  %``Atmospheric neutrino oscillation analysis with sub-leading effects in Super-Kamiokande I, II, and III,''
  Phys.\ Rev.\ D {\bf 81}, 092004 (2010)
  [arXiv:1002.3471 [hep-ex]].
  %%CITATION = ARXIV:1002.3471;%%
  %232 citations counted in INSPIRE as of 06 Oct 2015

%\cite{Aulakh:1999cd}
\bibitem{Aulakh:1999cd}
  C.~S.~Aulakh, A.~Melfo, A.~Rasin and G.~Senjanovic,
  %``Seesaw and supersymmetry or exact R-parity,''
  Phys.\ Lett.\ B {\bf 459}, 557 (1999)
  doi:10.1016/S0370-2693(99)00708-X
  [hep-ph/9902409].
  %%CITATION = doi:10.1016/S0370-2693(99)00708-X;%%
  %95 citations counted in INSPIRE as of 05 Jan 2016


%\cite{Khalil:2007dr}
\bibitem{Khalil:2007dr}
  S.~Khalil and A.~Masiero,
  %``Radiative B-L symmetry breaking in supersymmetric models,''
  Phys.\ Lett.\ B {\bf 665}, 374 (2008)
  [arXiv:0710.3525 [hep-ph]].
  %%CITATION = ARXIV:0710.3525;%%
  %69 citations counted in INSPIRE as of 12 Nov 2015

\bibitem{MOLINA:2014uha}
  J.~E.~Camargo-Molina, B.~O'Leary, W.~Porod and F.~Staub,
  %``On the vacuum stability of SUSY models,''
  PoS EPS {\bf -HEP2013}, 265 (2013)
  [arXiv:1310.1260 [hep-ph]].
  %%CITATION = ARXIV:1310.1260;%%
  %4 citations counted in INSPIRE as of 13 Nov 2015

  %\cite{Khalil:2015wua}
\bibitem{Khalil:2015wua}
  S.~Khalil and C.~S.~Un,
  %``Muon Anomalous Magnetic Moment in SUSY B-L Model with Inverse Seesaw,''
  arXiv:1509.05391 [hep-ph].
  %%CITATION = ARXIV:1509.05391;%%

%\cite{Holdom:1985ag}
\bibitem{Holdom:1985ag}
  B.~Holdom,
  %``Two U(1)'s and Epsilon Charge Shifts,''
  Phys.\ Lett.\ B {\bf 166}, 196 (1986);
  %%CITATION = PHLTA,B166,196;%%
  %905 citations counted in INSPIRE as of 13 Nov 2015
  %\cite{Babu:1997st}
%\bibitem{Babu:1997st}
  K.~S.~Babu, C.~F.~Kolda and J.~March-Russell,
  %``Implications of generalized Z - Z-prime mixing,''
  Phys.\ Rev.\ D {\bf 57}, 6788 (1998)
  [hep-ph/9710441];
  %%CITATION = HEP-PH/9710441;%%
  %161 citations counted in INSPIRE as of 15 Nov 2015
%\cite{delAguila:1988jz}
%\bibitem{delAguila:1988jz}
  F.~del Aguila, G.~D.~Coughlan and M.~Quiros,
  %``Gauge Coupling Renormalization With Several U(1) Factors,''
  Nucl.\ Phys.\ B {\bf 307}, 633 (1988)
  [Nucl.\ Phys.\ B {\bf 312}, 751 (1989)];
  %%CITATION = NUPHA,B307,633;%%
  %95 citations counted in INSPIRE as of 15 Nov 2015
%\cite{delAguila:1987st}
%\bibitem{delAguila:1987st}
  F.~del Aguila, J.~A.~Gonzalez and M.~Quiros,
  %``Renormalization Group Analysis of Extended Electroweak Models From the Heterotic String,''
  Nucl.\ Phys.\ B {\bf 307}, 571 (1988);
  %%CITATION = NUPHA,B307,571;%%
  %19 citations counted in INSPIRE as of 15 Nov 2015
  %\cite{Foot:1991kb}
%\bibitem{Foot:1991kb}
  R.~Foot and X.~G.~He,
  %``Comment on Z Z-prime mixing in extended gauge theories,''
  Phys.\ Lett.\ B {\bf 267}, 509 (1991);
  %%CITATION = PHLTA,B267,509;%%
  %142 citations counted in INSPIRE as of 15 Nov 2015
%\cite{Matsuoka:1986ig}
%\bibitem{Matsuoka:1986ig}
  T.~Matsuoka and D.~Suematsu,
  %``Low-energy Gauge Interactions From the $E(8)$ X $E(8)$-prime Superstring Theory,''
  Prog.\ Theor.\ Phys.\  {\bf 76}, 901 (1986).

%\cite{O'Leary:2011yq}
\bibitem{O'Leary:2011yq}
  B.~O'Leary, W.~Porod and F.~Staub,
  %``Mass spectrum of the minimal SUSY B-L model,''
  JHEP {\bf 1205}, 042 (2012)
  [arXiv:1112.4600 [hep-ph]].
  %%CITATION = ARXIV:1112.4600;%%
  %40 citations counted in INSPIRE as of 15 Nov 2015
  %\cite{Chankowski:2006jk}
\bibitem{Chankowski:2006jk}
  P.~H.~Chankowski, S.~Pokorski and J.~Wagner,
  %``Z-prime and the Appelquist-Carrazzone decoupling,''
  Eur.\ Phys.\ J.\ C {\bf 47}, 187 (2006)
  [hep-ph/0601097].
  %%CITATION = HEP-PH/0601097;%%
  %27 citations counted in INSPIRE as of 15 Nov 2015

%\cite{Abbas:2007ag}
%\bibitem{Abbas:2007ag}
  M.~Abbas and S.~Khalil,
  %``Neutrino masses, mixing and leptogenesis in TeV scale $B$ - L extension of the standard model,''
  JHEP {\bf 0804}, 056 (2008)
  doi:10.1088/1126-6708/2008/04/056
  [arXiv:0707.0841 [hep-ph]].
  %%CITATION = doi:10.1088/1126-6708/2008/04/056;%%
  %25 citations counted in INSPIRE as of 09 Jan 2016

%\cite{Abbas:2007ag}
\bibitem{Abbas:2007ag}
  M.~Abbas and S.~Khalil,
  %``Neutrino masses, mixing and leptogenesis in TeV scale $B$ - L extension of the standard model,''
  JHEP {\bf 0804}, 056 (2008)
  [arXiv:0707.0841 [hep-ph]].
  %%CITATION = ARXIV:0707.0841;%%
  %25 citations counted in INSPIRE as of 15 Nov 2015

%\cite{Mohapatra:1986bd}
\bibitem{Mohapatra:1986bd}
  R.~N.~Mohapatra and J.~W.~F.~Valle,
  %``Neutrino Mass and Baryon Number Nonconservation in Superstring Models,''
  Phys.\ Rev.\ D {\bf 34}, 1642 (1986);
  %%CITATION = PHRVA,D34,1642;%%
  %687 citations counted in INSPIRE as of 15 Nov 2015
%\cite{GonzalezGarcia:1988rw}
%\bibitem{GonzalezGarcia:1988rw}
  M.~C.~Gonzalez-Garcia and J.~W.~F.~Valle,
  %``Fast Decaying Neutrinos and Observable Flavor Violation in a New Class of Majoron Models,''
  Phys.\ Lett.\ B {\bf 216}, 360 (1989);
  %%CITATION = PHLTA,B216,360;%%
  %220 citations counted in INSPIRE as of 15 Nov 2015
%\cite{Khalil:2010iu}
%\bibitem{Khalil:2010iu}
  S.~Khalil,
  %``TeV-scale gauged B-L symmetry with inverse seesaw mechanism,''
  Phys.\ Rev.\ D {\bf 82}, 077702 (2010)
%  doi:10.1103/PhysRevD.82.077702
  [arXiv:1004.0013 [hep-ph]].
  %%CITATION = doi:10.1103/PhysRevD.82.077702;%%
  %36 citations counted in INSPIRE as of 17 Nov 2015
%\cite{Elsayed:2011de}
\bibitem{Elsayed:2011de}
  A.~Elsayed, S.~Khalil and S.~Moretti,
  %``Higgs Mass Corrections in the SUSY B-L Model with Inverse Seesaw,''
  Phys.\ Lett.\ B {\bf 715}, 208 (2012)
%  doi:10.1016/j.physletb.2012.07.066
  [arXiv:1106.2130 [hep-ph]];
  %31 citations counted in INSPIRE as of 17 Nov 2015
%\cite{Khalil:2010iu}
%\bibitem{Khalil:2010iu}
  S.~Khalil,
  %``TeV-scale gauged B-L symmetry with inverse seesaw mechanism,''
  Phys.\ Rev.\ D {\bf 82}, 077702 (2010)
  [arXiv:1004.0013 [hep-ph]].
  %%CITATION = ARXIV:1004.0013;%%
  %36 citations counted in INSPIRE as of 06 Oct 2015

%\cite{Porod:2003um}
\bibitem{Porod:2003um}
  W.~Porod,
  %``SPheno, a program for calculating supersymmetric spectra, SUSY particle decays and SUSY particle production at e+ e- colliders,''
  Comput.\ Phys.\ Commun.\  {\bf 153}, 275 (2003)
%  doi:10.1016/S0010-4655(03)00222-4
  [hep-ph/0301101];
  %%CITATION = doi:10.1016/S0010-4655(03)00222-4;%%
  %593 citations counted in INSPIRE as of 20 Nov 2015
%\cite{Porod:2011nf}
%\bibitem{Porod:2011nf}
  W.~Porod and F.~Staub,
  %``SPheno 3.1: Extensions including flavour, CP-phases and models beyond the MSSM,''
  Comput.\ Phys.\ Commun.\  {\bf 183}, 2458 (2012)
  doi:10.1016/j.cpc.2012.05.021
  [arXiv:1104.1573 [hep-ph]].
  %%CITATION = doi:10.1016/j.cpc.2012.05.021;%%
  %233 citations counted in INSPIRE as of 20 Nov 2015

%\cite{Staub:2008uz}
\bibitem{Staub:2008uz}
  F.~Staub,
  %``Sarah,''
  arXiv:0806.0538 [hep-ph];
  %%CITATION = ARXIV:0806.0538;%%
  %135 citations counted in INSPIRE as of 20 Nov 2015
 %\cite{Staub:2010jh}
%\bibitem{Staub:2010jh}
  F.~Staub,
  %``Automatic Calculation of supersymmetric Renormalization Group Equations and Self Energies,''
  Comput.\ Phys.\ Commun.\  {\bf 182}, 808 (2011)
%  doi:10.1016/j.cpc.2010.11.030
  [arXiv:1002.0840 [hep-ph]].
  %%CITATION = doi:10.1016/j.cpc.2010.11.030;%%
  %136 citations counted in INSPIRE as of 20 Nov 2015

%\cite{Hisano:1992jj}
\bibitem{Hisano:1992jj}
  J.~Hisano, H.~Murayama and T.~Yanagida,
  %``Nucleon decay in the minimal supersymmetric SU(5) grand unification,''
  Nucl.\ Phys.\ B {\bf 402}, 46 (1993)
%  doi:10.1016/0550-3213(93)90636-4
  [hep-ph/9207279];
  %%CITATION = doi:10.1016/0550-3213(93)90636-4;%%
  %396 citations counted in INSPIRE as of 20 Nov 2015
%\cite{Yamada:1992kv}
%\bibitem{Yamada:1992kv}
  Y.~Yamada,
  %``SUSY and GUT threshold effects in SUSY SU(5) models,''
  Z.\ Phys.\ C {\bf 60}, 83 (1993);
%  doi:10.1007/BF01650433
  %%CITATION = doi:10.1007/BF01650433;%%
  %97 citations counted in INSPIRE as of 20 Nov 2015
%\cite{Chkareuli:1998wi}
%\bibitem{Chkareuli:1998wi}
  J.~L.~Chkareuli and I.~G.~Gogoladze,
  %``Unification picture in minimal supersymmetric SU(5) model with string remnants,''
  Phys.\ Rev.\ D {\bf 58}, 055011 (1998)
%  doi:10.1103/PhysRevD.58.055011
  [hep-ph/9803335].
  %%CITATION = doi:10.1103/PhysRevD.58.055011;%%
  %66 citations counted in INSPIRE as of 20 Nov 2015

\bibitem{Ibanez:1982fr}
L.~E. Ibanez and G.~G. Ross,
% {\it $SU(2)_L \times U(1)$ symmetry breaking as a
 % radiative effect of supersymmetry breaking in GUTs},
 { Phys. Lett.} {\bf B110} (1982) 215;
%\bibitem{Inoue:1982pi}
K.~Inoue, A.~Kakuto, H.~Komatsu and S.~Takeshita,
 %``Aspects Of Grand Unified Models With Softly Broken Supersymmetry,''
 { Prog. Theor. Phys.} {\bf 68}, 927 (1982)
 [Erratum-ibid.\  {\bf 70}, 330 (1983)];
%\bibitem{Ibanez:1982ee}
L.~E. Ibanez,
%{\it Locally supersymmetric SU(5) grand unification},
{ Phys.  Lett.} {\bf B118} (1982) 73;
 J.~R. Ellis, D.~V. Nanopoulos,
and K.~Tamvakis,
%{\it Grand unification in simple supergravity},
  { Phys. Lett.} {\bf B121} (1983) 123;
%\bibitem{AlvarezGaume:1983gj}
L.~Alvarez-Gaume, J.~Polchinski, and M.~B. Wise,
%{\it Minimal low-energy supergravity},
{ Nucl. Phys.} {\bf B221} (1983) 495.

%\cite{Group:2009ad}
\bibitem{Group:2009ad}
  T.~E.~W.~Group [CDF and D0 Collaborations],
  %``Combination of CDF and D0 Results on the Mass of the Top Quark,''
  arXiv:0903.2503 [hep-ex].
  %%CITATION = ARXIV:0903.2503;%%
  %278 citations counted in INSPIRE as of 20 Nov 2015

%\cite{Gogoladze:2011db}
\bibitem{Gogoladze:2011db}
  I.~Gogoladze, R.~Khalid, S.~Raza and Q.~Shafi,
  %``Higgs and Sparticle Spectroscopy with Gauge-Yukawa Unification,''
  JHEP {\bf 1106}, 117 (2011)
  doi:10.1007/JHEP06(2011)117
  [arXiv:1102.0013 [hep-ph]].
  %%CITATION = doi:10.1007/JHEP06(2011)117;%%
  %32 citations counted in INSPIRE as of 20 Nov 2015

  %\cite{Gogoladze:2011aa}
\bibitem{Gogoladze:2011aa}
  I.~Gogoladze, Q.~Shafi and C.~S.~Un,
  %``Higgs Boson Mass from t-b-$\tau$ Yukawa Unification,''
  JHEP {\bf 1208}, 028 (2012)
  doi:10.1007/JHEP08(2012)028
  [arXiv:1112.2206 [hep-ph]];
  %%CITATION = doi:10.1007/JHEP08(2012)028;%%
  %53 citations counted in INSPIRE as of 20 Nov 2015
  %\cite{Ajaib:2013zha}
%\bibitem{Ajaib:2013zha}
  M.~Adeel Ajaib, I.~Gogoladze, Q.~Shafi and C.~S.~Un,
  %``A Predictive Yukawa Unified SO(10) Model: Higgs and Sparticle Masses,''
  JHEP {\bf 1307}, 139 (2013)
  doi:10.1007/JHEP07(2013)139
  [arXiv:1303.6964 [hep-ph]].
  %%CITATION = doi:10.1007/JHEP07(2013)139;%%
  %24 citations counted in INSPIRE as of 20 Nov 2015

%\cite{Belanger:2009ti}
\bibitem{Belanger:2009ti}
  G.~Belanger, F.~Boudjema, A.~Pukhov and R.~K.~Singh,
  %``Constraining the MSSM with universal gaugino masses and implication for
  %searches at the LHC,''
  JHEP {\bf 0911}, 026 (2009);
  %[arXiv:0906.5048 [hep-ph]].
  %%CITATION = JHEPA,0911,026;%%
H.~Baer, S.~Kraml, S.~Sekmen and H.~Summy,
  %``Dark matter allowed scenarios for Yukawa-unified SO(10) SUSY GUTs,''
  JHEP {\bf 0803}, 056 (2008).
  %[arXiv:0801.1831 [hep-ph]].
  %%CITATION = JHEPA,0803,056;%%

%\cite{Agashe:2014kda}
\bibitem{Agashe:2014kda}
  K.~A.~Olive {\it et al.} [Particle Data Group Collaboration],
  %``Review of Particle Physics,''
  Chin.\ Phys.\ C {\bf 38}, 090001 (2014).
  doi:10.1088/1674-1137/38/9/090001
  %%CITATION = doi:10.1088/1674-1137/38/9/090001;%%
  %2340 citations counted in INSPIRE as of 20 Nov 2015

%\cite{Amhis:2012bh}
\bibitem{Amhis:2012bh}
  Y.~Amhis {\it et al.}  [Heavy Flavor Averaging Group Collaboration],
  %``Averages of B-Hadron, C-Hadron, and tau-lepton properties as of early 2012,''
  arXiv:1207.1158 [hep-ex].
  %%CITATION = ARXIV:1207.1158;%%
  %142 citations counted in INSPIRE as of 25 Apr 2013

%\cite{Asner:2010qj}
\bibitem{Asner:2010qj}
  D.~Asner {\it et al.}  [Heavy Flavor Averaging Group Collaboration],
  %``Averages of b-hadron, c-hadron, and $\tau-lepton Properties,''
  arXiv:1010.1589 [hep-ex].
  %%CITATION = ARXIV:1010.1589;%%
  %443 citations counted in INSPIRE as of 01 May 2013

%\cite{Davier:2010nc}
\bibitem{Davier:2010nc}
  M.~Davier, A.~Hoecker, B.~Malaescu and Z.~Zhang,
  %``Reevaluation of the Hadronic Contributions to the Muon g-2 and to alpha(MZ),''
  Eur.\ Phys.\ J.\ C {\bf 71}, 1515 (2011)
  [Eur.\ Phys.\ J.\ C {\bf 72}, 1874 (2012)]
  [arXiv:1010.4180 [hep-ph]];
  %\cite{Hagiwara:2011af}
%\bibitem{Hagiwara:2011af}
  K.~Hagiwara, R.~Liao, A.~D.~Martin, D.~Nomura and T.~Teubner,
  %``(g-2)_mu and alpha(M_Z^2) re-evaluated using new precise data,''
  J.\ Phys.\ G {\bf 38}, 085003 (2011)
  [arXiv:1105.3149 [hep-ph]].
  %%CITATION = ARXIV:1105.3149;%%
  %330 citations counted in INSPIRE as of 29 juil. 2015
  %%CITATION = ARXIV:1010.4180;%%
  %430 citations counted in INSPIRE as of 29 juil. 2015

 %\cite{Baer:2012by}
\bibitem{Baer:2012by} See for instance;

  H.~Baer, I.~Gogoladze, A.~Mustafayev, S.~Raza and Q.~Shafi,
  %``Sparticle mass spectra from SU(5) SUSY GUT models with $b-\tau$ Yukawa coupling unification,''
  JHEP {\bf 1203}, 047 (2012)
  doi:10.1007/JHEP03(2012)047
  [arXiv:1201.4412 [hep-ph]];
  %%CITATION = doi:10.1007/JHEP03(2012)047;%%
  %13 citations counted in INSPIRE as of 20 Nov 2015
 %\cite{Li:2014xqa}
%\bibitem{Li:2014xqa}
  T.~Li, D.~V.~Nanopoulos, S.~Raza and X.~C.~Wang,
  %``A Realistic Intersecting D6-Brane Model after the First LHC Run,''
  JHEP {\bf 1408}, 128 (2014)
  doi:10.1007/JHEP08(2014)128
  [arXiv:1406.5574 [hep-ph]].
  %%CITATION = doi:10.1007/JHEP08(2014)128;%%
  %2 citations counted in INSPIRE as of 20 Nov 2015

%\cite{Flanz:1994yx}
\bibitem{Flanz:1994yx}
  M.~Flanz, E.~A.~Paschos and U.~Sarkar,
  %``Baryogenesis from a lepton asymmetric universe,''
  Phys.\ Lett.\ B {\bf 345}, 248 (1995)
  [Phys.\ Lett.\ B {\bf 382}, 447 (1996)]
  doi:10.1016/0370-2693(94)01555-Q
  [hep-ph/9411366];
  %%CITATION = doi:10.1016/0370-2693(94)01555-Q;%%
  %410 citations counted in INSPIRE as of 09 Jan 2016
%\cite{Flanz:1996fb}
%\bibitem{Flanz:1996fb}
  M.~Flanz, E.~A.~Paschos, U.~Sarkar and J.~Weiss,
  %``Baryogenesis through mixing of heavy Majorana neutrinos,''
  Phys.\ Lett.\ B {\bf 389}, 693 (1996)
  doi:10.1016/S0370-2693(96)01337-8
  [hep-ph/9607310];
  %%CITATION = doi:10.1016/S0370-2693(96)01337-8;%%
  %289 citations counted in INSPIRE as of 09 Jan 2016
  %\cite{Covi:1996wh}
%\bibitem{Covi:1996wh}
  L.~Covi, E.~Roulet and F.~Vissani,
  %``CP violating decays in leptogenesis scenarios,''
  Phys.\ Lett.\ B {\bf 384}, 169 (1996)
  doi:10.1016/0370-2693(96)00817-9
  [hep-ph/9605319];
  %%CITATION = doi:10.1016/0370-2693(96)00817-9;%%
  %637 citations counted in INSPIRE as of 09 Jan 2016
%\cite{Pilaftsis:1997jf}
%\bibitem{Pilaftsis:1997jf}
  A.~Pilaftsis,
  %``CP violation and baryogenesis due to heavy Majorana neutrinos,''
  Phys.\ Rev.\ D {\bf 56}, 5431 (1997)
  doi:10.1103/PhysRevD.56.5431
  [hep-ph/9707235];
  %%CITATION = doi:10.1103/PhysRevD.56.5431;%%
  %470 citations counted in INSPIRE as of 09 Jan 2016


%\cite{Grossman:2003jv}
\bibitem{Grossman:2003jv}
  Y.~Grossman, T.~Kashti, Y.~Nir and E.~Roulet,
  %``Leptogenesis from supersymmetry breaking,''
  Phys.\ Rev.\ Lett.\  {\bf 91}, 251801 (2003)
  doi:10.1103/PhysRevLett.91.251801
  [hep-ph/0307081];
  %%CITATION = doi:10.1103/PhysRevLett.91.251801;%%
  %118 citations counted in INSPIRE as of 09 Jan 2016
 %\cite{Chun:2004eq}
%\bibitem{Chun:2004eq}
  E.~J.~Chun,
  %``Late leptogenesis from radiative soft terms,''
  Phys.\ Rev.\ D {\bf 69}, 117303 (2004)
  doi:10.1103/PhysRevD.69.117303
  [hep-ph/0404029];
  %%CITATION = doi:10.1103/PhysRevD.69.117303;%%
  %23 citations counted in INSPIRE as of 09 Jan 2016
 %\cite{Grossman:2004dz}
%\bibitem{Grossman:2004dz}
  Y.~Grossman, T.~Kashti, Y.~Nir and E.~Roulet,
  %``New ways to soft leptogenesis,''
  JHEP {\bf 0411}, 080 (2004)
  doi:10.1088/1126-6708/2004/11/080
  [hep-ph/0407063];
  %%CITATION = doi:10.1088/1126-6708/2004/11/080;%%
  %65 citations counted in INSPIRE as of 09 Jan 2016
%\cite{Grossman:2005yi}
%\bibitem{Grossman:2005yi}
  Y.~Grossman, R.~Kitano and H.~Murayama,
  %``Natural soft leptogenesis,''
  JHEP {\bf 0506}, 058 (2005)
  doi:10.1088/1126-6708/2005/06/058
  [hep-ph/0504160];
  %%CITATION = doi:10.1088/1126-6708/2005/06/058;%%
  %30 citations counted in INSPIRE as of 09 Jan 2016
 %\cite{Kajiyama:2009ae}
%\bibitem{Kajiyama:2009ae}
  Y.~Kajiyama, S.~Khalil and M.~Raidal,
  %``Electron EDM and soft leptogenesis in supersymmetric B-L extension of the standard model,''
  Nucl.\ Phys.\ B {\bf 820}, 75 (2009)
  doi:10.1016/j.nuclphysb.2009.05.011
  [arXiv:0902.4405 [hep-ph]].
  %%CITATION = doi:10.1016/j.nuclphysb.2009.05.011;%%
  %6 citations counted in INSPIRE as of 09 Jan 2016

%\cite{Calibbi:2014pza}
\bibitem{Calibbi:2014pza}
  L.~Calibbi, A.~Mariotti, C.~Petersson and D.~Redigolo,
  %``Selectron NLSP in Gauge Mediation,''
  JHEP {\bf 1409}, 133 (2014)
  doi:10.1007/JHEP09(2014)133
  [arXiv:1405.4859 [hep-ph]];
  %%CITATION = doi:10.1007/JHEP09(2014)133;%%
  %12 citations counted in INSPIRE as of 09 Jan 2016
 %\cite{Gogoladze:2015jua}
%\bibitem{Gogoladze:2015jua}
  I.~Gogoladze, Q.~Shafi and C.~S.~Ün,
  %``Reconciling the muon g−2 , a 125 GeV Higgs boson, and dark matter in gauge mediation models,''
  Phys.\ Rev.\ D {\bf 92}, no. 11, 115014 (2015)
  doi:10.1103/PhysRevD.92.115014
  [arXiv:1509.07906 [hep-ph]].
  %%CITATION = doi:10.1103/PhysRevD.92.115014;%%
  %1 citations counted in INSPIRE as of 09 Jan 2016

 %\cite{Djouadi:2005gi}
\bibitem{Djouadi:2005gi}
  A.~Djouadi,
  %``The Anatomy of electro-weak symmetry breaking. I: The Higgs boson in the standard model,''
  Phys.\ Rept.\  {\bf 457}, 1 (2008)
  doi:10.1016/j.physrep.2007.10.004
  [hep-ph/0503172].
  %%CITATION = doi:10.1016/j.physrep.2007.10.004;%%
  %1000 citations counted in INSPIRE as of 06 Jan 2016

 %\cite{Djouadi:2005gj}
\bibitem{Djouadi:2005gj}
  A.~Djouadi,
  %``The Anatomy of electro-weak symmetry breaking. II. The Higgs bosons in the minimal supersymmetric model,''
  Phys.\ Rept.\  {\bf 459}, 1 (2008)
  doi:10.1016/j.physrep.2007.10.005
  [hep-ph/0503173].
  %%CITATION = doi:10.1016/j.physrep.2007.10.005;%%
  %899 citations counted in INSPIRE as of 06 Jan 2016

 \end{thebibliography}
\end{document}